\documentclass[aps,prb,showpacs, twocolumn,superscriptaddress]{revtex4-1}
\usepackage{moreverb, amsmath, amsfonts}
\usepackage{caption}
\usepackage{graphicx}
\usepackage{placeins}
\usepackage{algorithm2e}
\usepackage{algorithmic}
\usepackage{booktabs}
\usepackage{multirow}
\usepackage{xspace}
\usepackage{subfig}

\usepackage{adjustbox}
\usepackage{tikz}
\usetikzlibrary{patterns,calc}
\usepackage{pgf}

\newcommand{\BibTeX}{{\rm B\kern-.05em{\sc i\kern-.025em b}\kern-.08em
    T\kern-.1667em\lower.7ex\hbox{E}\kern-.125emX}}

\newcommand{\Or}{\mathcal{O}}
\renewcommand{\Im}{\mathrm{Im}~}
\newcommand{\Tr}{\mathrm{Tr}}

\newcommand{\JS}{\ensuremath{\mathcal J}\xspace}
\newcommand{\IS}{\ensuremath{\mathcal I}\xspace}
\newcommand{\RS}{\ensuremath{\mathcal R}\xspace}

\newcommand{\xc}{{\mathrm{xc}}}

\newcommand{\tot}{\mathrm{tot}}

\newcommand{\mc}[1]{\mathcal{#1}}

\newcommand{\ie}{\textit{i.e.}{}}
\newcommand{\eg}{\textit{e.g.}{}}

\newcommand{\abs}[1]{\left\lvert#1\right\rvert}

\newcommand{\averageop}[3]{\langle#1\lvert#2\rvert#3\rangle}

\newcommand{\braket}[2]{\langle#1\vert#2\rangle}

\newcommand{\ud}{\,\mathrm{d}}

\renewcommand{\Im}{\mathfrak{Im}}

\newcommand{\review}[1]{{#1}}

\begin{document}

\title{SIESTA-PEXSI: Massively parallel method for efficient and
accurate \textit{ab initio} materials simulation without matrix diagonalization} 
\author{Lin Lin}
\affiliation{Computational Research Division, Lawrence Berkeley National
Laboratory, Berkeley, CA 94720, USA}

\author{Alberto Garc\'{\i}a}
\affiliation{Institut de Ci\`encia de Materials de Barcelona,
  (ICMAB-CSIC), Campus de la
  UAB, E-08193 Bellaterra, Spain.}

\author{Georg Huhs}
\affiliation{
Barcelona Supercomputing Center,
Computer Applications in Science \& Engineering,
Edifici Nexus, Campus Nord UPC, c/ Gran Capit\`{a}, 2-4, \mbox{08034 Barcelona}, Spain.
}

\author{Chao Yang}
\affiliation{Computational Research Division, Lawrence Berkeley National
Laboratory, Berkeley, CA 94720, USA}

\begin{abstract}
  We describe a scheme for efficient large-scale electronic-structure calculations based on the combination of the pole expansion and selected inversion (PEXSI) technique with the SIESTA method, which uses numerical atomic orbitals within the Kohn-Sham density functional theory (KSDFT) framework.  The PEXSI technique can efficiently utilize the sparsity pattern of the Hamiltonian and overlap matrices generated in SIESTA, and for large systems has a much lower computational complexity than that associated with the matrix diagonalization procedure. The PEXSI technique can be used to evaluate the electron density, free energy, atomic forces, density of states and local density of states without computing any eigenvalue or eigenvector of the Kohn-Sham Hamiltonian. It can achieve accuracy fully comparable to that obtained from a matrix diagonalization procedure for general systems, including metallic systems at low temperature.  The PEXSI method is also highly scalable. With the recently developed massively parallel PEXSI technique, we can make efficient use of more than $10,000$ processors on high performance machines.  We demonstrate the performance and accuracy of the SIESTA-PEXSI method using several examples of large scale electronic structure calculations, including 1D, 2D and bulk problems with insulating, semi-metallic, and metallic character.
\end{abstract}


\maketitle

\section{Introduction}\label{sec:intro}
Kohn-Sham density functional theory (KSDFT) is the most widely used
framework for electronic-structure calculations, and plays an
important role in the analysis of electronic, structural and optical
properties of molecules, solids and nano-structures. The
efficiency of KSDFT depends largely on the computational
cost associated with the evaluation of the electron charge density for
a given potential within a self-consistent field (SCF) iteration.  The
most straightforward way to perform such an evaluation is to partially
diagonalize the Kohn-Sham Hamiltonian by computing a set of
eigenvectors corresponding to the algebraically smallest eigenvalues
of the Hamiltonian. The complexity of this approach is
$\mathcal{O}(N_e^3)$, where $N_e$ is the number of electrons in the
atomistic system of interest.  As the number of atoms or electrons in
the system increases, the cost of diagonalization becomes
prohibitively expensive.

Although linear scaling
algorithms~\cite{BowlerMiyazakiGillan2002,FattebertBernholc2000,HineHaynesMostofiEtAl2009,Yang1991,LiNunesVanderbilt1993,McWeeny1960,Goedecker1999,BowlerMiyazaki2012}
are attractive alternatives for improving the efficiency of KSDFT, they
rely on using the nearsightedness
principle~\cite{Kohn1996,ProdanKohn2005}, which asserts that the density
perturbation induced by a local change in the external potential decays
away from where the perturbation is applied.  One can then truncate elements of the density matrix
away from the diagonal.  Such truncation can be in practice applied only
to insulating systems whose density matrix elements decay exponentially
away from the diagonal, but not to metallic systems at low temperature,
for which the density matrix elements decay only algebraically away from
the diagonal.

The recently developed pole expansion and selected inversion (PEXSI)
method~\cite{LinLuYingE2009,LinLuYingEtAl2009,LinYangMezaEtAl2011,LinChenYangEtAl2013,JacquelinLinYang2014}
provides an alternative way for solving the Kohn-Sham problem
without using a diagonalization procedure, and without invoking
the nearsightedness principle to truncate density matrix elements. 
Compared to existing techniques, the PEXSI method has a few salient 
features: 1) 
PEXSI expresses physical quantities such as electron density, 
free energy, atomic forces, density of states and local density of 
states in terms of a spectral projector which is evaluated without
computing any eigenvalues or eigenvectors.
2)  The computational cost of
the PEXSI technique scales at most as $\Or(N_{e}^2$). The actual
computational cost depends on the dimensionality of the system: the cost
for quasi-1D systems such as nanotubes is $\Or(N_{e})$ \ie{}  linear
scaling; for quasi-2D systems such as graphene and surfaces (slabs) is
$\Or(N_{e}^{1.5})$; for general 3D bulk systems is $\Or(N_{e}^2)$.  
\review{This favorable scaling hinges on the sparse
character of the Hamiltonian and overlap matrices, but not on any
fundamental assumption about the localization properties
 of the single particle density matrix.}
3) The PEXSI
technique can be accurately applied to general materials system
including small gapped systems and metallic systems, and remains
accurate at low temperatures.  4) The PEXSI method has a two-level
parallelism structure and is by design highly scalable.
The recently developed massively parallel PEXSI technique can make
efficient usage of $10,000\sim 100,000$ processors on high performance
machines.  5) PEXSI can be controlled with a few input
parameters, and can act nearly as a black-box substitution of the
diagonalization procedure commonly used in electronic structure
calculations.

In order to benefit from the PEXSI method, the Hamiltonian and overlap
matrices must be sparse.  This requirement is satisfied if
localized discretization is used for representing the Kohn-Sham
Hamiltonian by a finite sized matrix.  Examples of localized
discretization include numerical atomic
orbitals~\cite{Junquera:01,MohanChen2010,KennyHorsfieldFujitani2000,Ozaki:03,BlumGehrkeHankeEtAl2009},
Gaussian type
orbitals~\cite{FrischPopleBinkley1984,VandeVondeleKrackMohamedEtAl2005},
the finite difference~\cite{ChelikowskyTroullierSaad1994} and finite
element~\cite{TsuchidaTsukada1995} methods, adaptive curvilinear
coordinates~\cite{TsuchidaTsukada1998}, optimized nonorthogonal
orbitals~\cite{BowlerMiyazakiGillan2002,FattebertBernholc2000,HineHaynesMostofiEtAl2009}
and adaptive local basis functions~\cite{LinLuYingE2012}.  In contrast,
the plane-wave basis set is not localized and therefore
cannot directly benefit from the PEXSI method.  Even though they are
formally localized, the
number of degrees of freedom per atom associated with methods such as 
the finite difference and the finite element is usually much
larger than that associated with other methods such as numerical
atomic orbitals, leading to an increase of the preconstant factor in the
computational cost.  Therefore the finite difference and the
finite element methods may not benefit as much from the PEXSI technique
as those based on numerical atomic orbitals. We note also that the
use of hybrid functionals with an orbital-based exact exchange
term~\cite{Becke1993,AdamoBarone1999} may significantly impact the
sparsity pattern of the Hamiltonian matrix and increase the
computational cost.

In previous work~\cite{LinChenYangEtAl2013}, the applicability of
the PEXSI method was demonstrated for accelerating atomic orbital based
electronic structure calculations. With the sequential implementation of
the PEXSI method, it was possible to perform electronic structure
calculations accurately for a nanotube containing 10,000 atoms
discretized by a single-$\zeta$ (SZ, minimal) basis, and to perform geometry
optimization for a nanotube that contains more than $1000$ atoms with
a double-$\zeta$ plus polarization (DZP) basis. However, the sequential
implementation  does not benefit from the inherent parallelism in the
PEXSI method, and therefore leads to limited or no improvement for
general electronic structure calculations.

\review{The contribution of this paper is twofold: 1) We present
the SIESTA-PEXSI
method, which combines the SIESTA method~\cite{Soler2002,Artacho2008}
based on numerical atomic orbitals and the recently developed massively
parallel PEXSI method~\cite{JacquelinLinYang2014}.}
The SIESTA-PEXSI method can be efficiently
scalable to more than $10,000$ processors. We provide performance data
for a range of systems, including strong and weak scaling characteristics,
and illustrate the crossover points beyond which the new approach is more
efficient than diagonalization. The accuracy of the result obtained from
the SIESTA-PEXSI method is nearly indistinguishable from the result
obtained from the diagonalization method.  
\review{2) We develop a hybrid scheme of density of states estimation and
Newton's method to obtain the chemical potential.}
We demonstrate that the scheme is highly efficient and robust with
respect to the initial guess, with or without the presence of gap
states.  The SIESTA-PEXSI approach has been implemented as a new solver
in SIESTA, with built-in heuristics that balance efficiency and
accuracy, but at the same time offering full control by the user.

This paper is organized as follows.  In section~\ref{sec:theory}, we
describe the massively parallel PEXSI technique, and how to integrate it
with the SIESTA method. We also present a new method to update the
chemical potential.  In section~\ref{sec:numerical}, we report the
performance of the SIESTA-PEXSI method on several problems.

Throughout the paper, we use $\Im(A)$ to denote the imaginary part of a
complex matrix $A$.  We use $H, S$ to denote the discretized
Hamiltonian matrix and the corresponding overlap matrix obtained
from a basis set $\Phi$ such as numerical atomic orbitals.
Similarly $\hat{\gamma}(x,x')$ denotes the single particle density
matrix operator, and the corresponding electron density
is denoted by ${\rho}(x)$.  The matrix $\Gamma$ denotes the 
single particle density matrix represented in the $\Phi$ basis.
In PEXSI, $\Gamma$ and related matrices are approximated by a finite
$P$-term pole expansion, denoted by
$\Gamma_{P},\Gamma^{\mc{F}}_{P},\Gamma^{E}_{P}$ respectively.  However,
to simplify notation, we will drop the subscript $P$ and simply use
$\Gamma,\Gamma^{\mc{F}},\Gamma^{E}$ to denote the approximated matrices
unless otherwise noted.

\section{Theory and Practical Implementation}\label{sec:theory}

\subsection{Basic formulation}\label{subsec:basic}

The ground-state electron charge density ${\rho}(x)$ of an atomistic
system can be obtained from the self-consistent solution to
the Kohn-Sham equations
\begin{equation}
  \hat{H}\left[{\rho}(x)\right] \psi_i(x) = \psi_i(x) \varepsilon_i,
\label{kseqs}
\end{equation}
where $\hat{H}$ is the Kohn-Sham Hamiltonian that depends on ${\rho}(x)$,
$\{\psi_i(x)\}$ are the Kohn-Sham orbitals which in turn determine the
charge density by
\begin{equation}
  {\rho}(x) = \sum_i^{\infty} |\psi_i(x)|^2 f_i
\label{rhodef}
\end{equation}
with occupation numbers $f_i$ that can be chosen according to the 
Fermi-Dirac distribution function
\begin{equation}
f_i= f_{\beta} (\varepsilon_i - \mu) = \frac{2}{1+e^{\beta(\varepsilon_i-\mu)}},
\label{fermidirac}
\end{equation}
where $\mu$ is the chemical potential chosen to ensure that
\begin{equation}
  \int {\rho}(x) dx = N_e,
\label{chargesum1}
\end{equation}
and $\beta$ is the inverse of the temperature, i.e.,
$\beta = 1/(k_B T)$ with $k_B$ being the Boltzmann constant.

The most straightforward method to solve the Kohn-Sham problem is to
expand the orbitals $\psi_i$ as a linear combination of a finite
number of basis functions $\{\varphi_j\}$, and thus recast \eqref{kseqs} as a
(generalized) eigenvalue problem within an iterative procedure to
achieve self-consistency in the charge density. The computational
complexity of this approach is $\Or(N^3)$, where $N$ is the number of
basis functions and is generally proportional to the number of
electrons $N_e$ or atoms in the system to be studied. This approach
becomes prohibitively expensive when the size of the system increases.

Formally, the electronic-structure problem can be recast in terms of
the one-particle density matrix defined by
\begin{equation}
\hat{\gamma}=\sum_{i=1}^{\infty}
|\psi_{i}\rangle f_{\beta}(\varepsilon_i-\mu) 
\langle \psi_i| = f_{\beta} (\hat{H} - \mu),
\label{gammaeq}
\end{equation}
%
%
%
with $\mu$ chosen so that $\mbox{Tr}\left[\hat{\gamma}\right] = N_e$.
$\hat{\gamma}$ can thus be evaluated without the need for
diagonalization, if the Fermi function is approximated by a linear
combination of a number of simpler functions. This is the Fermi
operator expansion (FOE) method~\cite{Goedecker1993}. 

While most of the FOE schemes require as many as $\Or(\beta\Delta E)$
or $\Or(\sqrt{\beta\Delta E})$ terms of simple functions (with $\Delta
E$ the spectrum width), the recently
developed pole expansion~\cite{LinLuYingE2009} is particularly
promising since it requires only $\Or(\log \beta\Delta E)$ terms of
simple rational functions. The pole expansion has the analytic
expression
\begin{equation}
  f_{\beta}(\varepsilon-\mu) \approx \Im \sum_{l=1}^{P}
  \frac{\omega^{\rho}_l}{\varepsilon-(z_l+\mu)}, \label{eqn:polerho}
\end{equation}
We refer readers to
Ref.~\onlinecite{LinLuYingE2009,LinChenYangEtAl2013} for more details.
The complex shifts $\{z_{l}\}$ and weights $\{\omega^{\rho}_l\}$ are
determined only by $\beta,\Delta E$ and the number of poles $P$. All
quantities in the pole expansion are known explicitly and their
calculation takes negligible amount of time.

Following the derivation in Ref.~\onlinecite{LinChenYangEtAl2013}, we
can use (\ref{eqn:polerho}) to approximate the single particle density
matrix $\hat{\gamma}$ by its $P$-term pole expansion, denoted by
$\hat{\gamma}_{P}$ as

\begin{equation}
  \begin{split}
    \hat{\gamma}_{P}(x,x') &= \Phi(x) \Im\left(
    \sum_{l=1}^{P}\frac{\omega^{\rho}_l}{H - (z_l+\mu) S}\right)
    \Phi^T(x')\\
    &\equiv \Phi(x) \Gamma \Phi^T(x').
  \end{split}
  \label{eqn:gammapole}
\end{equation}
where $\Phi=[\varphi_1,\cdots,\varphi_{N}]$ is a collective vector of
basis functions, $S_{ij}=\braket{\varphi_{i}}{\varphi_{j}}$,
$H_{ij}=\averageop{\varphi_{i}}{\hat{H}}{\varphi_{j}}$, and $\Gamma$
is an $N\times N$ matrix represented in terms of the 
$\Phi$ basis.  To simplify our notation, we will drop the subscript $P$
originating from the $P$-term pole expansion approximation unless
otherwise noted.

It would seem that the need to carry out $P$ matrix inversions in
(\ref{eqn:gammapole}) would mean that the computational complexity of
this approach is still close to the $\Or(N^3)$ scaling of
diagonalization. However, what is really needed in practice is just
the electron density in real space, that is
\begin{equation}
    {\rho}(x) = \Phi(x) \Gamma \Phi^T(x)
     = \sum_{ij}\Gamma_{ij}\varphi_{j}(x)\varphi_{i}(x).
  \label{eqn:approxrho}
\end{equation}
When the basis functions $\varphi_{i}(x)$ are compactly supported in
real space, the product of two functions $\varphi_i(x)$ and
$\varphi_j(x)$ will be zero when they do not overlap. These $i,j$
pairs can be excluded from the summation in
Eq.~\eqref{eqn:approxrho}. Consequently, we {\em only} need
$\Gamma_{ij}$ such that $\varphi_{j}(x)\varphi_{i}(x)\ne 0$ in
Eq.~\eqref{eqn:approxrho}. This set of $\Gamma_{ij}$'s is a subset of
$\{\Gamma_{ij}\vert H_{ij}\ne 0\}$. To obtain these {\em selected
  elements}, we need to compute the corresponding elements of $(H -
(z_l+\mu)S)^{-1}$ for all $z_l$. 
\review{We emphasize that we compute the selected elements of the
density matrix because only these elements are needed to compute
physical quantities such as charge density, energy and forces, due to
the localized character of the basis set.  The computed selected
elements of the density matrix are accurate, and should be regarded as
if we performed a conventional $\Or(N^3)$ calculation first, and then
only kept the corresponding selected elements of the density matrix.  In
principle we could retrieve any matrix element of the density matrix,
simply by enlarging the set of ``selected elements''.  This process is
fundamentally different from the usage of ``near-sightedness''
approximation, which throws away the information of the density matrix
beyond the truncation region.}

The recently developed selected inversion
method~\cite{LinLuYingEtAl2009,LinYangMezaEtAl2011}
provides an efficient way of computing the selected elements of an
inverse matrix. For a (complex) symmetric matrix of the form $A=H-zS$,
the selected inversion algorithm first constructs an $LDL^T$
factorization of $A$, where $L$ is a block lower triangular matrix
called the Cholesky factor, and $D$ is a block diagonal matrix. The
computational scaling of the selected inversion algorithm is only
proportional to the number of nonzero elements in the Cholesky factor
$L$,  which is $\Or(N)$ for quasi-1D systems, $\Or(N^{1.5})$ for 
quasi-2D systems,
and $\Or(N^{2})$ for 3D bulk systems, thus achieving universal
asymptotic improvement over the diagonalization method for systems of all
dimensions.  It should be noted that the selected inversion algorithm is
an \textit{exact} method for computing selected elements of $A^{-1}$
if exact arithmetic is employed, and in practice the only source of
error originates from the roundoff error.  In particular, the selected
inversion algorithm does not rely on any localization property of
$A^{-1}$.

In addition to computing the charge density at a reduced computational
complexity in each SCF iteration, we can also use this pole-expansion
and selected-inversion (PEXSI) technique to compute the free energy
and the atomic forces efficiently without diagonalizing the Kohn-Sham
Hamiltonian. Following the derivation in
Ref.~\onlinecite{LinChenYangEtAl2013}, 
the relevant expressions are:
\begin{equation}
  \begin{split}
  \mc{F}_{\tot} =&  \Tr[\Gamma^{\mc{F}} S] + \mu N_e
  - \frac12 \iint \frac{{\rho}(x)
  {\rho}(y)}{\abs{x-y}} \ud x \ud y\\
  &
  + E_{\xc}[{\rho}]
  - \int V_{\xc}[{\rho}] {\rho}(x) \ud x,
  \end{split}
  \label{eqn:Helmholtzpole}
\end{equation}
\begin{equation}
	F_{I} = -\frac{\partial \mc{F}_{\tot}}{\partial R_I} =
  -\Tr\left[ \Gamma \frac{\partial H}{\partial R_I} \right]
      +\Tr\left[ \Gamma^E \frac{\partial S}{\partial R_I} \right].
  \label{eqn:forcepole}
\end{equation}
where the energy and free-energy density matrices $\Gamma^{E}$
and $\Gamma^{\mc{F}}$ are given by pole expansions with \textit{the
  same poles} as those used for computing the charge density:
\begin{equation}
  \Gamma^{E,\mc{F}} = \Im \sum_{l=1}^{P}
  \frac{\omega^{E,\mc{F}}_l}{H - (z_l+\mu) S}.
  \label{eqn:gammaF}
\end{equation}
Since the trace terms in~\eqref{eqn:Helmholtzpole}
and~\eqref{eqn:forcepole} require only the $(i,j)$th entries of
$\Gamma^{E,\mc{F}}$ for $(i,j)$ satisfying $S_{ij} \neq 0$ or $H_{ij}
\neq 0$, the needed elements of the energy-density matrices can be
computed without additional complexity. 

\subsection{Massively parallel PEXSI method}\label{subsec:ppexsi}

In addition to its favorable asymptotic complexity, the
PEXSI method is also inherently more scalable than the standard
approach based on matrix diagonalization when it is implemented on 
a parallel computer.  The parallelism in PEXSI exists at two levels.
First, the selected inversions associated with different poles 
(usually on the order of $40\sim 60$) 
are completely independent.  
Second, each selected inversion itself can be parallelized by
using the parallel selected inversion method called
PSelInv~\cite{JacquelinLinYang2014}.

A parallel selected inversion consists of the following steps:
\begin{enumerate}
\item The rows and columns of the matrices $H$ and $S$ are reordered 
to reduce the number of nonzeros in the triangular factor of the $LDL^T$ 
decomposition of $H - z S$. 
\item A parallel symbolic factorization of $H - z S$ is performed to 
identify the location of the nonzero matrix elements in $L$.
\item The numerical $LDL^T$ decomposition (or equivalent $LU$
  decomposition) of $H - zS$ is performed.
\item The desired selected elements of $(H-z S)^{-1}$ are computed from
      $L$ and $D$.
\end{enumerate}

Step 1 can be performed in parallel by using the
ParMETIS~\cite{KarypisKumar1998a} or the
PT-Scotch~\cite{ChevalierPellegrini2008} software packages. Its cost is
much smaller compared to the numerical factorization, and only needs to
be done once per SCF cycle.  Although for symmetric matrices only
$LDL^{T}$ factorization is needed, the PEXSI package currently use the
SuperLU\_DIST software package~\cite{LiDemmel2003} to perform steps 2
and 3 in parallel with $LU$ factorization. The $LU$ factorization
contains equivalent information as in $LDL^{T}$ factorization but
can be twice as expensive in the worst case due to the lack of usage of
the symmetry property of the matrix.  
PEXSI and PSelInv has a independent data structure, and allows to be
interfaced with other sparse direct solvers.

The cost of symbolic factorization in Step 2 is usually much lower
than the numerical factorization.  The numerical factorization
procedure can be described in terms of the traversal of a tree called
the
{\em elimination tree}.  Each node of the tree corresponds to a block of
continuous columns of $H -
zS$. A node $\RS$ is the parent of a node $\JS$ if and only if
\begin{equation}
	\RS = \min\left\{ \IS > \JS ~ | ~ L_{\IS,\JS} \mbox{ is a nonzero block}
	\right\}.
	\label{eq:parent}
\end{equation}
In SuperLU\_DIST, each node is distributed among a subset of processors
block cyclically.  The traversal of the elimination tree
proceeds from the leaves towards the root. The update of $\IS$ is
performed in parallel on a subset of processors assigned to $\IS$ and
its children nodes, and the main operations involved in the update are a
number of dense matrix-matrix multiplications. In addition to
parallelism within the update of a supernode, additional concurrency can
be exploited in the traversal of different branches of the elimination
tree for updating different supernodes. 

All of these techniques can be used in Step 4 to compute
selected elements of $(H-zS)^{-1}$. In this step, 
the elimination 
tree is traversed from the root down towards the leaves.  As each node 
is traversed, selected elements of $(H-zS)^{-1}$ within 
the columns that are mapped to that node are computed through 
a number of dense matrix-matrix multiplications. Communication is
needed among processors that are mapped to the node and its ancestors.
Multiple nodes belonging to different branches of the elimination 
tree can be traversed simultaneously if the update of 
these nodes do not involve communications with the same ancestor.

For sufficiently large problems, SuperLU\_DIST can achieve substantial 
speedup in the numerical factorization when hundreds to thousands of
processors are used.  Similar or better speedup factors can be observed
when selected elements of $(H-zS)^{-1}$ are computed from the
distributed $L$ and $D$ factors.  It is shown that for large matrices 
a single selected inversion can scale to $4,096$ cores, and if $40$ poles are
used in the pole expansion of the Fermi-Dirac function, then total
number of computational cores that can be efficiently utilized by the
parallel PEXSI method is $40\times 4096 \approx 160,000$.  
We will give some more concrete examples in section~\ref{sec:numerical}
to demonstrate the performance of our implementation of the parallel
PEXSI algorithm. 

  
Compared to PEXSI, it is generally difficult to efficiently use that
many computational cores in a dense matrix calculation algorithm such as
those implemented in ScaLAPACK. The main bottleneck of the computation
using ScaLAPACK
is the reduction of a
dense matrix to a tridiagonal matrix and the back transformation of the
eigenvectors of a tridiagonal matrix, which are inherently sequential.
Moreover, the cost of the diagonalization method scales cubically with
respect to the matrix dimension. This limits the size of the matrix that
can be handled by ScaLAPACK, as well as the parallel scalability on
massively parallel computers.
Although some recent progress has been made to make these
transformations more efficient~\cite{Auckenthaler2011}, making diagonalization
scalable on more than tens of thousands of cores remains
a challenging task.

\subsection{Determination of the chemical potential}
\label{subsec:chemicalpotential}

The chemical potential $\mu$ required in the pole expansion
\eqref{eqn:gammapole} is not known a priori. Its value must in general
be determined so that the total number of electrons is appropriate:
\begin{equation}
N_e = N_{\beta}(\mu) = \Tr[\hat{\gamma}] = \Tr[\Gamma \Phi^T \Phi] = \Tr[\Gamma S].
\label{eqn:Nmu}
\end{equation}
As the right-hand-side is a non-decreasing function of $\mu$, the
chemical potential can be efficiently obtained by Newton's
method, maybe combined with extra safeguards such as bisection. The
required derivative $N'_{\beta}(\mu)$ can be computed with very little
extra cost using the pole expansion of the derivative of the Fermi-Dirac
distribution $f'_{\beta}(\varepsilon -
\mu)$, which can be constructed, for the reasons illustrated in
Ref.~\onlinecite{LinChenYangEtAl2013}, by using the same shifts $z_l$ as
those in \eqref{eqn:polerho}:
\begin{equation}
	N'_{\beta}(\mu) = \Tr[\Gamma^{d} S],
	\label{}
\end{equation}
\begin{equation}
	\Gamma^{d} =  \Im \sum_{l=1}^{P}
  \frac{\omega^{d}_l}{H - (z_l+\mu) S}.
	\label{}
\end{equation}

When Newton's method is used, the convergence of $\mu$ is rapid near
the correct chemical potential. However, the standard Newton's method
may not be robust enough when the initial guess is far away from the
correct chemical potential. It may give, for example, too large a
correction when $N_{\beta}'(\mu)$ is close to zero, as when $\mu$ is
near the edge or in the middle of a band gap.


One way to overcome the above difficulty is to use an approximation to
the function $N_\beta(\varepsilon)$ to narrow down the region in which
the correct $\mu$ must lie. This function can be seen effectively as a
(temperature smeared) cumulative density of states, counting the
number of eigenvalues in the interval $(-\infty,\varepsilon)$. 
We can evaluate $N_{\infty}(\varepsilon)$, its zero-temperature limit,
without computing any eigenvalues of $(H,S)$.  Instead, we perform a
matrix decomposition of the shifted matrix $H-\varepsilon S = LDL^T$,
where $L$ is unit lower triangular and $D$ is diagonal.  It follows
from Sylvester's law of inertia~\cite{Sylvester1852}, which states
that the inertia (the number of negative, zero and positive
eigenvalues) of a real symmetric matrix does not change under a
congruent transform, that $D$ has the same inertia as that of
$H-\varepsilon S$. Hence, we can obtain $N_{\infty}(\varepsilon)$ by simply
counting the number of negative entries in $D$.  Note that the matrix
decomposition $H-\varepsilon S = L D L^{T}$ can be computed
efficiently by using
a sparse $LDL^{T}$ or $LU$ factorization in real arithmetic.  It
requires fewer floating point operations than the complex arithmetic
direct sparse factorization used in PEXSI.

\begin{figure}[h]
  \begin{center}
    \subfloat[]{\includegraphics[width=0.5\linewidth]{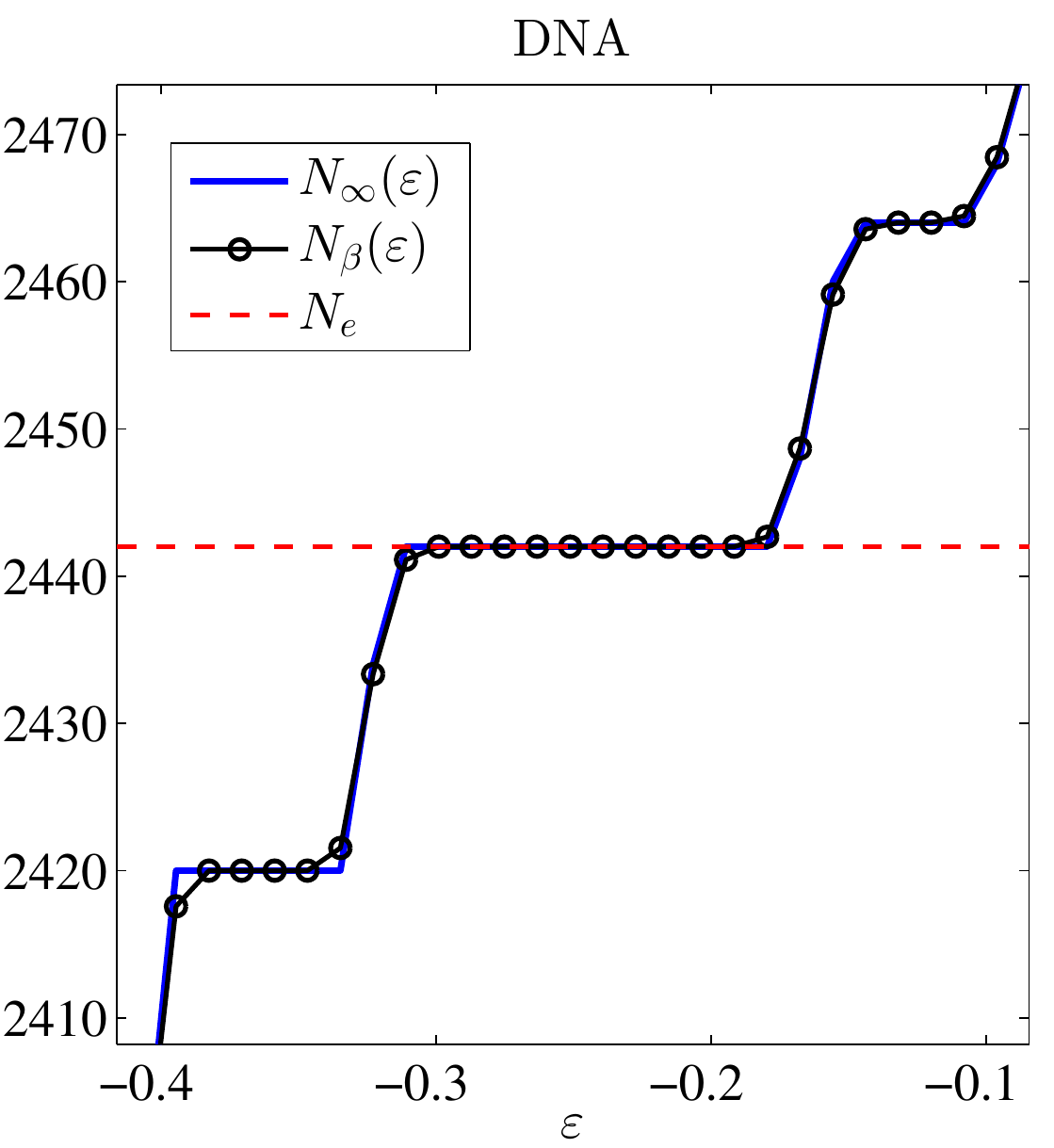}}
    \subfloat[]{\includegraphics[width=0.5\linewidth]{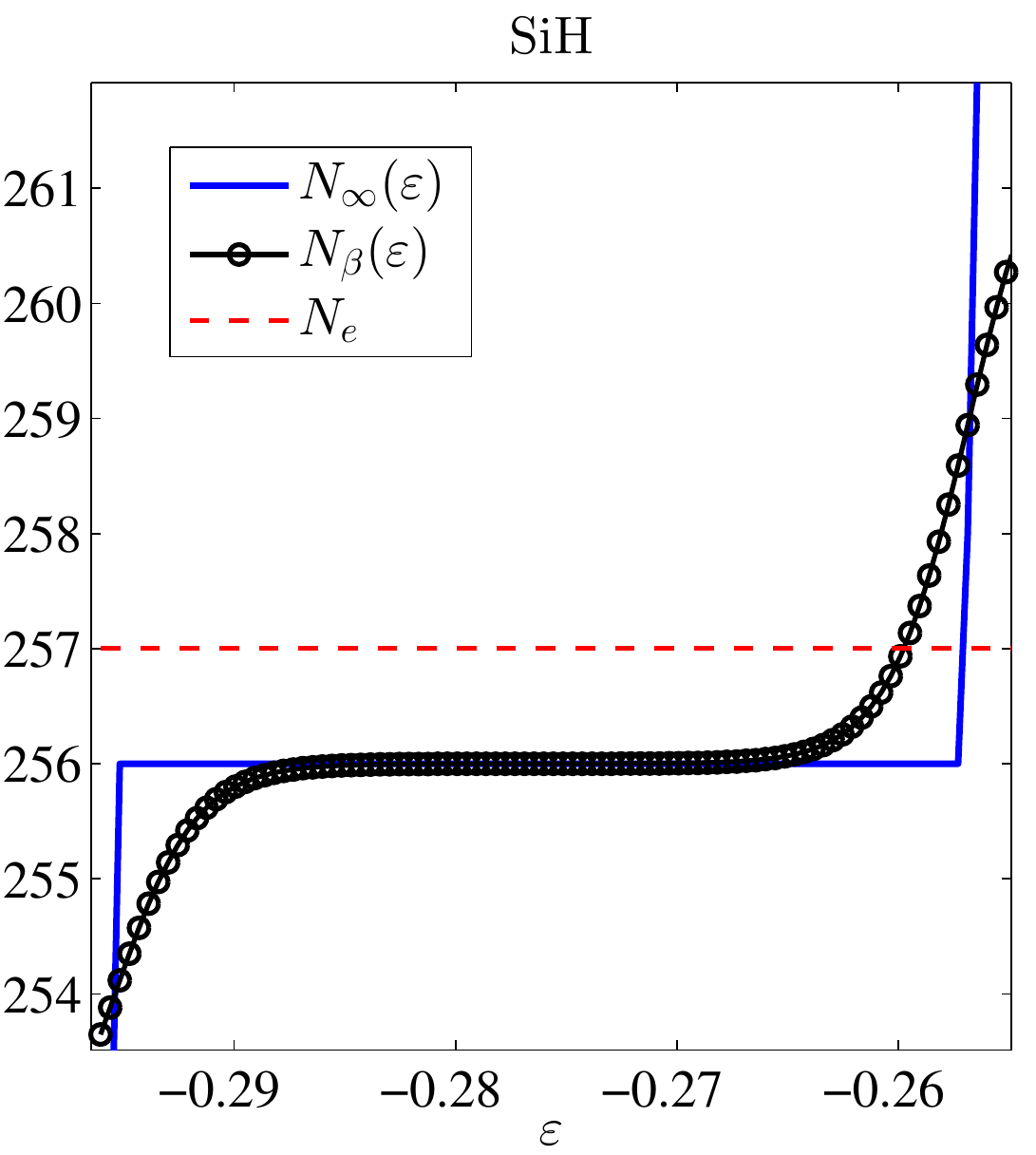}}
  \end{center}
  \caption{(color online) Number of electrons at zero temperature
  $N_{\infty}(\varepsilon)$ (blue solid line), at 300 K
  $N_{\beta}(\varepsilon)$ (black line with circles), and exact number
  of electrons $N_{e}$ (red dashed line) for (a) DNA-1 (b) SiH. (unit of
  $\varepsilon$: Hartree) 
  } 
  \label{fig:Ne}
\end{figure}

To estimate $N_{\beta}(\mu)$ for a finite $\beta$, we use 
the identity
\begin{equation}
	N_{\beta}(\mu) 
	= \int_{-\infty}^{\infty} f_{\beta}(\varepsilon-\mu)
	\ud N_{\infty}(\varepsilon),
\end{equation}
and perform an integration by parts to obtain
\begin{equation}
	N_{\beta}(\mu) 
	= -\int_{-\infty}^{\infty} f'_{\beta}(\varepsilon-\mu)
	N_{\infty}(\varepsilon) \ud \varepsilon,
	\label{eqn:finiteelectronnumber}
\end{equation}
The integral in \eqref{eqn:finiteelectronnumber} can be
evaluated numerically by sampling $f'_{\beta}(\varepsilon-\mu)$ and
$N_{\infty}(\varepsilon)$ at a number of quadrature points
$\{\varepsilon_{m}\}_{m=1}^{Q}$ and performing a weighted sum of
$f'_{\beta}(\varepsilon_{m}-\mu)N_{\infty}(\varepsilon_m)$ for
$m=1,...,Q$. The $Q$ evaluations of
$\{N_{\infty}(\varepsilon_{m})\}_{m=1}^{Q}$ can be performed
simultaneously using the $LDL^T$ factorization-based inertia counting
procedure described above, with $Q$ groups of processors. Since the
derivative of the Fermi-Dirac function is sharply peaked, we can
approximate $N_{\beta}(\mu)$ in a given interval by sampling
$N_{\infty}(\varepsilon)$ in a slightly wider interval.
Fig.~\ref{fig:Ne} shows the number of electrons at zero temperature
$N_{\infty}(\varepsilon)$ obtained from inertia counting procedure, and the
interpolated finite temperature profile $N_{\beta}(\varepsilon)$ at
$300$ K for a DNA system (with finite gap) and a SiH system (with zero
gap) near the
Fermi energy.
While the finite temperature smearing effect is
negligible for insulators at $300$ K, it is more pronounced for metals
and leads to a more smooth $N_{\beta}(\varepsilon)$ which is suitable for applying Newton's method
to find the chemical potential.  On the other hand, the inertia counting
procedure obtains the global profile of the cumulative density of
states, and does not suffer from the problem of being trapped in
intermediate band gaps.

%
%

This approximation to the function $N_{\beta}(\varepsilon)$ is then
available for use in determining the approximate placement of the true
$\mu$ by simple root finding: $N_{\beta}(\mu)=N_e$. In practice, we
need to start from an interval which is large enough to contain the
chemical potential. For systems with a gap the interval should also
contain the gap edges. In this case, two roots are sought:
$N_{\beta}(\varepsilon_l)=N_e-\delta N$ and
$N_{\beta}(\varepsilon_h)=N_e+\delta N$, where $\delta N$ is small,
say 0.1, and so $\varepsilon_l$ and $\varepsilon_h$ will be estimates
of the band edges from which $\mu$ can be determined as
$\mu=1/2(\varepsilon_l+\varepsilon_h)$.

The fidelity of $N_{\beta}(\varepsilon)$, and thus the quality of the
estimation of $\mu$, depends on the density with which the interval can
be sampled, which with a fixed number $Q$ of sampling points for
$N_{\infty}(\varepsilon)$ will increase as the interval is narrowed. In
practice, the above procedure is repeated with progressively smaller
intervals until either the $\mu$ estimate stabilizes or the size of the
interval is small enough.  Unless the original interval is very large, 2
or 3  
inertia counts are enough to provide an adequate $\mu$ from which to
start the PEXSI process, and the number of inertia counts can be one or
even zero (i.e. the inertia counting procedure is turned off) 
in subsequent SCF iterations.
Newton's method in the solver then takes over 
for the final refining of $\mu$.

The hybrid procedure of inertia counting and Newton's method is
efficient and robust for both insulating and metallic systems.  This
will be further demonstrated in section~\ref{sec:numerical}.

\subsection{Calculation of density of states and localized density of
states}\label{subsec:dos}

The density of states $g(\varepsilon)$ is defined as
\begin{equation}
	g(\varepsilon) = 2 \sum_{i} \delta(\varepsilon - \varepsilon_i).
	\label{eqn:DOS}
\end{equation}
The factor of two comes from spin degeneracy. As advanced in the
previous section, the cumulative density of states (CDOS)
\begin{equation}
	C(\varepsilon) = \int_{-\infty}^{\varepsilon} g(\varepsilon') \ud \varepsilon'.
	\label{eqn:cdos}
\end{equation}
is exactly the function $N_{\infty}(\varepsilon)$ discussed there.  To
evaluate $g(\varepsilon)$, we sample $N_{\infty}(\varepsilon)$ at a set
of $\{\varepsilon_{l}\}$ using the inertia counting procedure, and use
an appropriate interpolation scheme to approximate
$N_{\infty}(\varepsilon)$ for other values of $\varepsilon$.  The
approximation to the DOS is then obtained by numerical
differentiation.  We remark that inertia counting is not the only
diagonalization-free method for computing the DOS.  Other methods that
make use of matrix vector multiplications only are also possible (see
\eg a recent review~\cite{LinSaadYang2013}).

Another physical quantity that can be easily approximated via selected 
inversion is the localized density of states (LDOS), defined as 
\begin{equation}
	g(x,\varepsilon) = 2 \sum_{i} \delta(\varepsilon-\varepsilon_{i})
	\abs{\psi_{i}(x)}^2.
	\label{eqn:LDOS}
\end{equation}
and representing in fact the contribution to the charge density of the
states with eigenvalues in the vicinity of $\varepsilon$, as filtered
by the $\delta$ function.

When a finite basis set $\Phi$ is used, the LDOS can be represented as
\begin{equation}
	g(x,\varepsilon) =  2 \Phi(x) C \delta(\varepsilon-\Xi) C^T \Phi^{T}(x).
	\label{eqn:LDOSbasis}
\end{equation}
where $C$ is a matrix of the coefficients of the expansion in $\Phi$ of the
orbitals $\psi_{i}(x)$, and $\Xi$ is a diagonal matrix of the
eigenvalues. It follows from the Sokhotski-Plemelj
formula~\cite{PlemeljRadok1964}
\begin{equation}
	\lim_{\eta \to 0+} \frac{1}{\varepsilon+i \eta} =
	\mathrm{PV}(1/\varepsilon) - i \pi
	\delta(\varepsilon),
	\label{eqn:Plemelj}
\end{equation}
that
\begin{equation}
	\delta(\varepsilon-\varepsilon_{i}) = \lim_{\eta\to 0+} -\frac{1}{\pi}
	\Im \frac{1}{\varepsilon+i\eta - \varepsilon_{i}},
	\label{eqn:ImPlemlj}
\end{equation}
where symbol $\mathrm{PV}$ in \eqref{eqn:Plemelj} stands for the 
Cauchy principal value.

Combining Eq.~\eqref{eqn:LDOSbasis} and~\eqref{eqn:ImPlemlj},
we obtain the following alternative expression for the LDOS: 
\begin{equation}
	\begin{split}
	g(x,\varepsilon) 
	&= -\frac{2}{\pi} \lim_{\eta\to 0+} 
	\Phi(x) C \left(\varepsilon+i\eta - \Xi \right)^{-1} C^T \Phi^{T}(x) \\
	&= \frac{2}{\pi} \lim_{\eta\to 0+}\Phi(x) \left[H-(\varepsilon+i\eta)S\right]^{-1}
	\Phi^{T}(x).
	\end{split}
	\label{eqn:LDOSgreen}
\end{equation}
Note that Eq.~\eqref{eqn:LDOSgreen} allows us to compute the LDOS without 
using any eigenvalue or eigenvector.
In practice, we take $\eta$ to be a small positive number, and the local
DOS can be approximated by
\begin{equation}
  g(x,\varepsilon)\approx \frac{2}{\pi} \sum_{ij}\varphi_{i}(x)\varphi_{j}(x) 
	\left[ H - (\varepsilon+i\eta) S\right]^{-1}_{ij},
	\label{eqn:LDOSGreen}
\end{equation}
which is similar to the computation of the electron density.  Again
only the selected elements of the matrix $\left[ H -
  (\varepsilon+i\eta) S\right]^{-1}$ are needed for the computation of
the LDOS for each $\varepsilon$, and the selected elements can be
obtained efficiently by the PSelInv procedure.

\subsection{Interface to SIESTA and heuristics for enhancing the efficiency}
\label{sec:heuristic}

SIESTA is a density functional theory code which uses finite-range atomic
orbitals to discretize the Kohn-Sham
problem~\cite{Soler2002}, and thus handles internally
sparse H, S, and single-particle density matrices. SIESTA is therefore
well suited to implement the PEXSI method along the lines explained
above.  

The PEXSI method can be directly integrated into SIESTA as a new kind
of electronic-structure solver. Conceptually, the interface between
SIESTA and PEXSI is straightforward. The existing SIESTA framework
takes care of setting up the basis set and of constructing the sparse
$H$ and $S$ matrices at each iteration of the self-consistent-field
(SCF) cycle. $H$ and $S$ are passed to the PEXSI module, which returns
the density matrix $\Gamma$ and, optionally, the energy-density matrix
$\Gamma^E$ (needed for the calculation of forces) and the
$\Gamma^{\mc{F}}$
matrix that can be used to estimate the electronic entropy. SIESTA
then computes the charge density to generate a new Hamiltonian to
continue the cycle, until convergence is achieved. Energies and forces
are computed as needed.

The details of the interface are controlled by a number of parameters
which provide flexibility to the user, especially in regard to the
bracketing and tolerances involved in the determination of the
chemical potential, and in the context of parallel computation. We
list some of the more relevant parameters in Table~\ref{tab:param},
but we should note that even finer control is possible by other, more
specialized parameters.

\begin{table}[htbp]
\begin{center}
  \begin{tabular}{|c|c|}
    \hline
    parameter & purpose  \\
    \hline
    $[\mu_{lb}, \mu_{ub}]$ & \begin{minipage}[t]{0.3\textwidth}Initial guess of the lower and upper bounds of the chemical potential $\mu$ \end{minipage} \\ \hline
    $n_{IC}$ & \begin{minipage}[t]{0.3\textwidth}The number of SCF steps in which inertial count is used 
               to narrow down the interval in which $\mu$ lies.\end{minipage} \\ \hline
    $tol_{NE}$ & Tolerance on number of electrons\\ \hline
    $T$  &  Electronic temperature \\ \hline
    $P$  &  Number of poles  \\ \hline
    $ppp$    & Number of processors used per pole  \\ \hline
    $np_0$     & \begin{minipage}[t]{0.3\textwidth}The number of processors used for non-PEXSI operations \end{minipage}\\ \hline
  \end{tabular}
\end{center}
\caption{Main SIESTA-PEXSI parameters}
\label{tab:param}
\end{table}

The initial interval $[\mu_{lb},\mu_{ub}]$ should be large enough to
contain the true chemical potential.  If it does not, the code will
automatically expand it until $\mu$ is properly bracketed, but it is
obviously more efficient to start the process with an appropriate
interval, even if it is relatively large, and use the program's refining
features without the need for backtracking. Note that, due to the
implicit reference energy used by SIESTA, $\mu$ is typically negative
and within a Rydberg of zero, so the specification of
$[\mu_{lb},\mu_{ub}]$ should not be a problem in practice even when
nothing is known about the electronic structure of the system. In
fact the program can choose an appropriately wide starting interval if the
user does not indicate one.

The bracketing interval can be refined by the use of the inertia
counting
procedure detailed in~\ref{subsec:chemicalpotential}. This is particularly
important in the first few SCF steps in which the approximate
electron density (and consequently the Hamiltonian) is far from
converged. The parameter $n_{IC}$ controls directly the number of SCF steps for
which this procedure is followed. Beyond those, the PEXSI solver is
directly invoked without further
refinement. There is also the possibility of linking the use of
inertia counting refinement to the convergence level of the
calculation. 

Because the chemical potential tends to oscillate in the first few SCF
steps, a completely robust method should, in principle, search for a
bracket afresh at every step, starting from a sufficiently wide
interval. However, it is wasteful to completely ignore the previous
bracketing when the SCF cycle reaches a more stable region, so we
developed a heuristic to allow SIESTA-PEXSI to reuse the bracket
determined in the previous SCF step, under various conditions related
to the convergence level and to whether or not inertia counting is still
used as a safeguard. We set the $\mu$ search interval
for a new SCF step to be slightly larger (by a value that can also be
controlled) than the final $\mu$ interval used in the previous SCF
step. For some systems, an estimation of the change in the
band-structure energy caused by the change in $H$ can be used to shift
the bracket across iterations. If $\mu$ ever falls out of the
bracketing interval, the algorithm recovers automatically by expanding
the interval appropriately.

When the search interval is deemed appropriate, we invoke the
PEXSI solver with a starting $\mu$ equal to the mid-point of the
interval, and use the solver's built-in Newton's method to refine
$\mu$ until the error in the total number of electrons is below
$tol_{NE}$. 

The tolerance in the computed number of electrons $tol_{NE}$ is a key
parameter of the PEXSI module, and it should be set to an
appropriately small value to guarantee the necessary accuracy in the
results. But there is no advantage to setting the tolerance too low in
the early stages of the SCF cycle. Hence, the favored mode of
operation of SIESTA-PEXSI is to use an on-the-fly tolerance level
which ranges from a coarse value at the start and progressively (in
tandem with the reduction of a typical convergence-level metric, which
can also control the coarseness of the bracketing) decreases towards
the desired fine level $tol_{NE}$.
Typically, no more than $3-5$ PEXSI solver iterations are needed to
achieve a high accuracy in $N_e$, and in practice the use of an
adaptive tolerance level with proper bracketing means that just one or
at most two solver iterations are enough for most of the steps in a
complete SCF cycle. For gapped systems, with the starting $\mu$ well
into the gap, the solver iteration can be turned off without affecting
the accuracy of the results, leading to significant savings.

The total cost of a SCF cycle includes the $LDL^T$ factorizations
required in the inertia counting bracket refinements, in addition to the
cost of the PEXSI solver. As the cost of an inertia counting invocation
is typically much lower than a solver iteration, it is clear that a
strategy that uses at least a few inertia counting steps to properly
bracket $\mu$ before invoking the solver will be in fact cheaper than
one which incurs the extra costs associated with a bad guess of $\mu$.
The overall cost of the algorithm would depend on the uniformity of
the convergence to self-consistency and on the electronic structure of
the system (i.e, whether or not it has a gap). We will report the
actual cost of $\mu$ search for a variety of systems in the next
section.

The design of SIESTA-PEXSI provides some flexibility to users in terms
of the usage of computational resources.  If the user's goal is a low
time-to-solution on a large machine, the number of processors per pole
\textit{ppp} should be increased as much as possible.  The overall
number of processors \textit{Np} should be set to
$P\times$\textit{ppp}, where $P$ is the number of poles, to achieve
complete parallelization over poles. If the goal is to minimize the
number of processors involved in the computation, then $ppp$ should be
set to the minimum number that allows the problem to fit in memory.
It should be noted that the memory requirements of the PEXSI approach
are significantly lower than those of the diagonalization-based
algorithm, as the relevant matrices are handled in their original
sparse form, and not converted to dense form as in ScaLAPACK. As the
parallel efficiency over the number of poles is nearly perfect, the total cost
(time$\times$\textit{Np}) does not depend on the total number of
processors used, so the minimum-cost strategy can be used with a
useful range of machine sizes: from total parallelization over poles
in medium-to-large machines, to serial calculation of poles in small
machines.

The number of processors that can be profitably used by the PEXSI
module is typically larger than the number of processors that the
non-solver part of SIESTA needs (as it is itself very efficient,
having been coded for essentially O(N) operation).  Hence the
non-solver operations in the SIESTA side use a subset of the
processors available, specified by the $np_0$ parameter in
Table~\ref{tab:param}. Appropriate logic is in place to orchestrate
the data movement and control flow.

%

\section{Numerical results}\label{sec:numerical}

In this section we report the performance and accuracy achieved 
by the massively parallel SIESTA-PEXSI method for computing
the ground state energy and atomic forces of several systems.
We also show some other capabilities of SIESTA-PEXSI that may be
useful for characterizing the electronic properties of materials.

To demonstrate that SIESTA-PEXSI can handle different types of
systems, we choose five different test problems for our numerical
experiments, including insulating, semi-metallic, and metallic systems,
and covering all the relevant dimensions. Table~\ref{tab:system}
gives a brief description of each system.

Our calculations were performed on the Edison system at the National
Energy Research Scientific Computing (NERSC) center. Each node
consists of two twelve-core Intel ``Ivy Bridge'' 2.4-GHz processors
and has 64 gigabytes (GB) of DDR3 1866-MHz memory.

\begin{table}
\begin{center}
  \begin{tabular}{|c|c|c|}\hline
 Name & type & Description \\
\hline
 DNA  & \begin{minipage}[t]{0.08\textwidth}1D insulating \end{minipage} & \begin{minipage}[t]{0.3\textwidth}
\begin{flushleft}
DNA. The basic unit (715 atoms) contains two base pairs of an A-DNA
double helix. Replicating this unit along the axis of rotation results
in several instances of this problem of different sizes, with
quasi-one-dimensional character. The largest instance considered
contains 25 units and is 76 nanometers long.
\end{flushleft}
\end{minipage}
\\ \hline
 C-BN & \begin{minipage}[t]{0.08\textwidth}2D semi-metallic\end{minipage} & \begin{minipage}[t]{0.3\textwidth}
\begin{flushleft}
A layer of boron nitride (BN) on top of a graphene sheet. Several
instances of this problem are generated by varying the orientation
of the BN sheet relative to the graphene layer within an appropriate
periodic cell. The largest example considered of this
quasi-two-dimensional system has 12700 atoms.
\end{flushleft}
\end{minipage}
\\ \hline
H$_2$O & \begin{minipage}[t]{0.08\textwidth}3D insulating\end{minipage} & \begin{minipage}[t]{0.3\textwidth}
\begin{flushleft}
Liquid water.  The basic repeating unit contains 64 molecules, and is
generated by taking a snapshot of a molecular-dynamics run with the
TIP4P force field. Appropriate supercells can be generated, with the
largest having 8000 molecules, or 24000 atoms.
\end{flushleft}
\end{minipage} \\ \hline
Al & \begin{minipage}[t]{0.08\textwidth}3D metallic\end{minipage} & \begin{minipage}[t]{0.3\textwidth}
\begin{flushleft}
A bulk Al system generated from a $8\times 8\times 8$ supercell of the
primitive $FCC$ unit cell. The positions of the atoms are perturbed by
small random displacements to break the symmetry. This is a typical
metallic system.
\end{flushleft}
\end{minipage} \\ \hline
SiH & \begin{minipage}[t]{0.08\textwidth}3D, special\end{minipage} & \begin{minipage}[t]{0.3\textwidth}
\begin{flushleft}
A bulk Si system with 64 atoms, with an H interstitial impurity. The Fermi
level in this system is pinned by the position of an H-derived level
within the gap. 
\end{flushleft}
\end{minipage} \\ \hline
  \end{tabular}
\end{center}
\caption{Test problems used in numerical experiments.}
\label{tab:system}
\end{table}

The number of atoms in the various instances of the problems listed on
Table~\ref{tab:system}, the sizes of unit cells, as well as the matrix
sizes and the sparsity (i.e., the percentage of nonzero elements) of
the corresponding $H$ matrices and their $L$ and $U$ factors are given in
Table~\ref{tab:exampleFeatures}. Note that the number of nonzero
elements in $L+U$ is the same as the number of nonzero elements in
$L+D+L^{T}$ if an equivalent $LDL^{T}$ factorization is to be used.
We use DNA-$a$ to denote a DNA strand
with $a$ unit cells, C-BN$_\alpha$ to denote C-BN layers in which
the Boron-Nitride layer is rotated by $\alpha$ degrees relative to the
graphene sheet (for this set of systems a lower $\alpha$ implies a
larger system size), 
and H$_2$O-$n$ to denote a box of liquid
water with $n$ unit cells.  The various instances of DNA, C-BN,
and H$_2$O are used to test the performance (including parallel
scaling) and accuracy of SIESTA-PEXSI, while the smaller Al and SiH
systems are used mainly to showcase the accuracy of SIESTA-PEXSI and
the effectiveness of the hybrid inertia counting plus Newton's method for
finding the correct chemical potential for metallic systems.

In all cases we use a DZP basis set, which results in 13 orbitals per
atom for C, N, B, O, and P, and 5 orbitals per H atom.

\begin{figure}[htb]
  \begin{center}
    \subfloat[]{\includegraphics[width=0.7\linewidth]{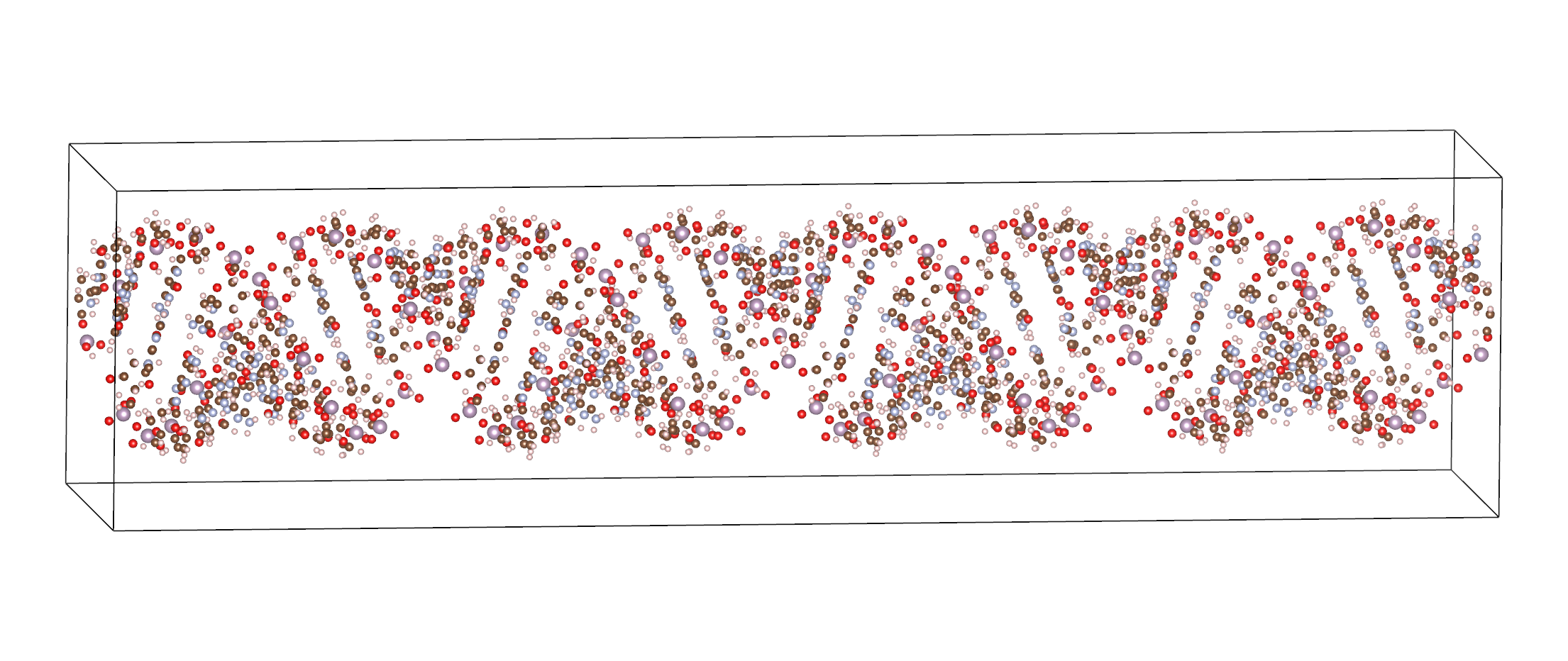}}

    \subfloat[]{\includegraphics[width=0.7\linewidth]{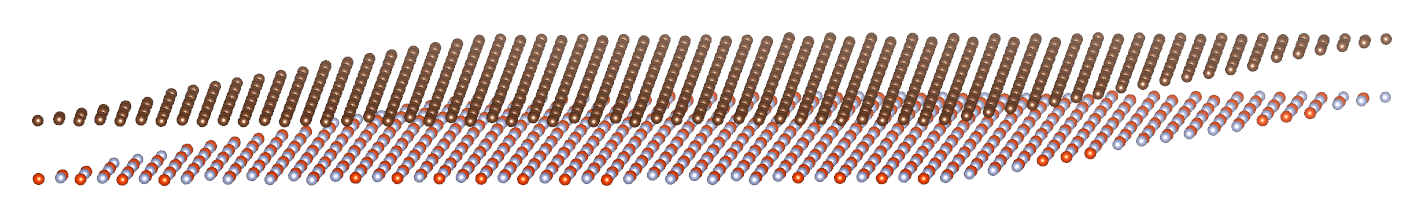}}

    \subfloat[]{\includegraphics[width=0.4\linewidth]{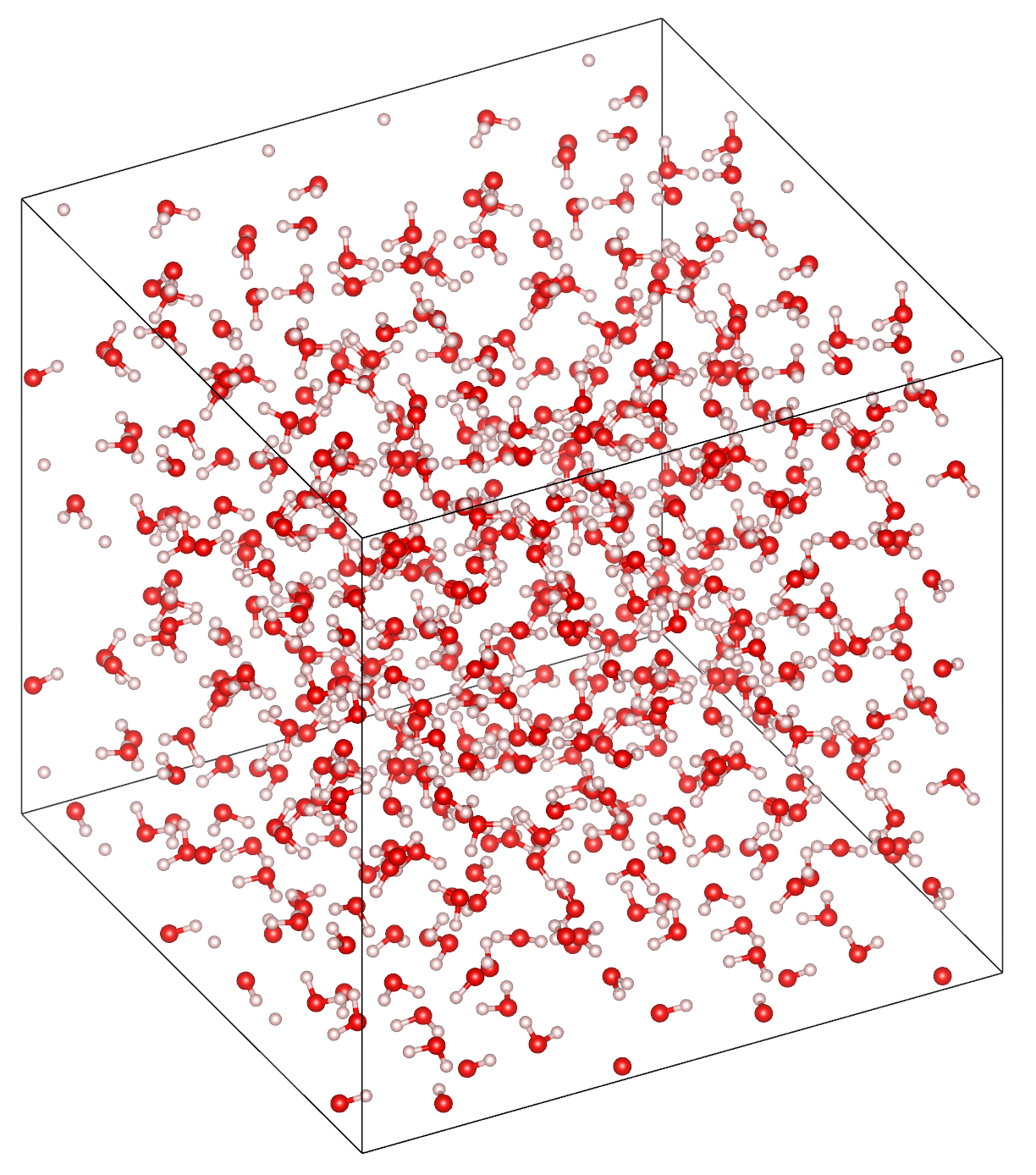}} 
  \end{center}
  \caption{(color online) The atomic configuration of (a) the DNA-4 system (b) the
  C-BN$_{2.3}$ system and (c) the H$_{2}$O-8 system.}
  \label{fig:examples}
\end{figure}


\begin{table}[htb]
\begin{center}
  \begin{tabular}{@{\extracolsep{3pt}}|l|rrrrr|}
    \hline
    Example & Atoms & $N$ & $s_H$ & $s_{LU}$ & $l$ (nm)  \\
    \hline
    DNA-1                &   715 &   7183 &  6.8\%  &  23\%  &  3.1 \\ 
    DNA-4                &  2860 &  28732 &  1.7\%  & 6.2\%  & 12 \\ 
    DNA-9                &  6435 &  64647 &  0.75\% & 2.9\%  & 28 \\ 
    DNA-16               & 11440 & 114928 &  0.42\% & 1.7\%  & 49 \\ 
    DNA-25               & 17875 & 179575 &  0.27\% & 1.1\%  & 76 \\ 
    \hline
    C-BN$_{2.3}$           &  1988 &   25844 &  5.9\%  & 36\%  & 5.5 \\ 
    C-BN$_{1.43}$          &  3874 &   50362 &  3.0\%  & 24\%  & 7.7 \\ 
    C-BN$_{0.57}$          &  7988 &  103844 &  1.5\%  & 15\%  & 11  \\ 
    C-BN$_{0.00}$          & 12770 &  166010 &  0.91\%  & 11\%  & 14 \\ 
    \hline
    H$_2$O-8                  &   1536 &   11776 &   2.3\%  &   28\%  & 2.5 \\ 
    H$_2$O-27                 &   5184 &   39744 &  0.69\%  &   18\%  & 3.7 \\ 
    H$_2$O-64                 &  12288 &   94208 &  0.29\%  &   12\%  & 5.0 \\ 
    H$_2$O-125                &  24000 &  184000 &  0.15\%  &  8.4\%  & 6.2 \\ 
    \hline
    Al                        &    512 &    6656 &    36\%  &  94\%  & 1.6 \\ %
    \hline
    SiH                       &     65 &     833 &    74\%  &  97\%  & 1.1 \\ 
    \hline
  \end{tabular}
\end{center}
\caption{Characteristics of the test examples in terms of the number of
atoms, physical dimension of each system, the corresponding size of the 
Hamiltonian $H$, and the percentage of nonzero elements in $H$ and in the 
$L$ and $U$ factors.  Also given is the
length scale $l$, which is in the case of DNA the length of the cell in
the direction of rotation axis of the strand, for C-BN the length of a
side of the unit cell,   and for H$_2$O the side length of the cube. 
}
\label{tab:exampleFeatures}
\end{table}

\subsection{Accuracy of the SIESTA-PEXSI approach}\label{subsec:accuracy}

We now report the accuracy of SIESTA-PEXSI in terms of the computed
energies and atomic forces.  We measure the accuracy of the energy by
examining the difference $err_E$ between the free energy computed by
SIESTA-PEXSI and that computed by the standard SIESTA approach in
which free energies are obtained from density matrices constructed
from the eigenvectors of $(H,S)$. A similar metric is used for
assessing the accuracy of atomic forces.  We denote by $err_F$ the
maximum force difference among all atoms. 

In Table~\ref{tab:err} we list the differences of energy per atom and
force for fully converged calculations for a number of test problems
with indication of the
number of poles used in the pole expansion and of the target tolerance
for electron count used in the chemical potential search.  
In all our tests we set the electronic temperature to
300K. 
As we can see from this table, the errors in the energy
computed by SIESTA-PEXSI are typically of the order of $10^{-6}$ eV
per atom. The maximum error in atomic forces is of the order of
$10^{-5}$ eV/\AA{}. Both are sufficiently small for most applications,
and can be achieved with a modest number of poles and reasonable
accuracy tolerance on $\mu$.  For SiH, $P=60$ and $tol_{ne}=10^{-4}$
bring the error in the energy per atom to the $\mu eV$
level. In this more delicate case, one needs the extra accuracy to
locate precisely the Fermi level, which is pinned in a gap state.

%
\begin{table}[htbp]
\begin{center}
\begin{tabular}{|c|c|c|c|c|}  
\hline
 System & $P$ & $tol_{ne}$ & $err_E$(eV/atom) & $err_F$(eV/\AA{})  \\
\hline
  DNA-1  & 50 & $10^{-4}$ &  $2\times 10^{-8}$ & $2\times 10^{-6}$  \\ \hline
  C-BN$_{2.3}$  & 40 & $10^{-3}$ & $3\times 10^{-6}$ & $2\times 10^{-5}$ \\ \hline
  H$_2$O-8 & 40 & $10^{-4}$ & $3\times 10^{-6}$ & $6 \times 10^{-5}$ \\ \hline
  Al     & 40 & $10^{-3}$ & $6\times 10^{-6}$ & $2 \times 10^{-6}$ \\ \hline
  SiH    & 40 & $10^{-3}$ & $10^{-4}$         & 7$ \times 10^{-5}$ \\ \hline
  SiH    & 60 & $10^{-4}$ & $6\times 10^{-6}$ & $7 \times 10^{-5}$ \\ \hline
\end{tabular}
\end{center}
\caption{Deviations of the energy and forces obtained from SIESTA-PEXSI
with respect to those from a diagonalization-based calculation.}
\label{tab:err}
\end{table}

\subsection{Efficiency of the SIESTA-PEXSI approach}\label{subsec:efficiency}

We now report the performance of SIESTA-PEXSI in comparison to that of
the standard SIESTA approach in which density matrices are obtained
from the eigenvectors of $(H,S)$ computed by the ScaLAPACK
diagonalization procedure based on the suite of subroutines {\tt
  pdpotrf}, {\tt pdsyngst}, {\tt pdsyevd/pdsyevx}, and {\tt pdtrsm}
for transformation to a standard eigenvalue problem, solution, and
back transformation, respectively. 

We use instances of the DNA, C-BN and H$_2$O problems to test both
the strong scaling of the solver, which is measured by the change in
wallclock time as a function of the number of processors used to solve
a problem of fixed size, and the weak scaling, which is measured by
the change of wall clock time as we increase in tandem both the
problem size and the number of processors used in the computation.

Our timing measurements refer to a single SCF step.  For completeness
they include also the time used to setup the Hamiltonian and the
overlap matrices, but this is in any case a very small fraction of the
total ($<5\%$) for these systems.  
The symbolic factorization is inexpensive compared to numerical
factorization and selected inversion, and can be computed once for the
entire SCF iteration using only the sparsity structure of $H$ and $S$,
and the time for
symbolic factorization is excluded in the timings.
It should be noted that the data for the diagonalization-based method
include the time involved in constructing the density and energy-density
matrices from the eigenvectors, needed at each SCF step.

The cost of the diagonalization method is determined directly by $H$ and
$S$ and the internal, basically hardwired operating parameters in
ScaLAPACK. The chemical potential is computed
from the list of eigenvalues. In contrast,
SIESTA-PEXSI determines the chemical potential iteratively in each SCF
step, and thus the computational load depends on the actual sequence of
bracketing and refining of $\mu$ followed. In order to provide an
appropriate reference with which to compare the diagonalization-based
results, our SIESTA-PEXSI timings include one (H$_2$O) or two (DNA and C-BN) 
inertia counting cycles to narrow down the search interval for the 
chemical potential and one call of the PEXSI solver. 

%

As mentioned above, one or two calls of the PEXSI
solver are typically needed for an appropriately precise computation
of $\mu$ as the SCF cycle unfolds. Given a good strategy for keeping a
tight bracketing of $\mu$, the final steps in the cycle close to
convergence are likely to need just one call, and the same behavior is
expected for most of the cycle when a good guess of the starting
electronic structure can be provided, as in molecular-dynamics or
geometry-optimization simulation. It should also be noted that later steps in
the SCF cycle will not need the inertia counting procedure.  Our timings
for SIESTA-PEXSI are thus representative on average of the
computational effort expected for a given system. In some cases, the
actual effort will be higher by a small factor, and in others (as in
systems with a gap) might even be smaller.

\begin{figure}[h]
  \centering
  \subfloat[DNA-25]{\includegraphics[width=0.8\linewidth]{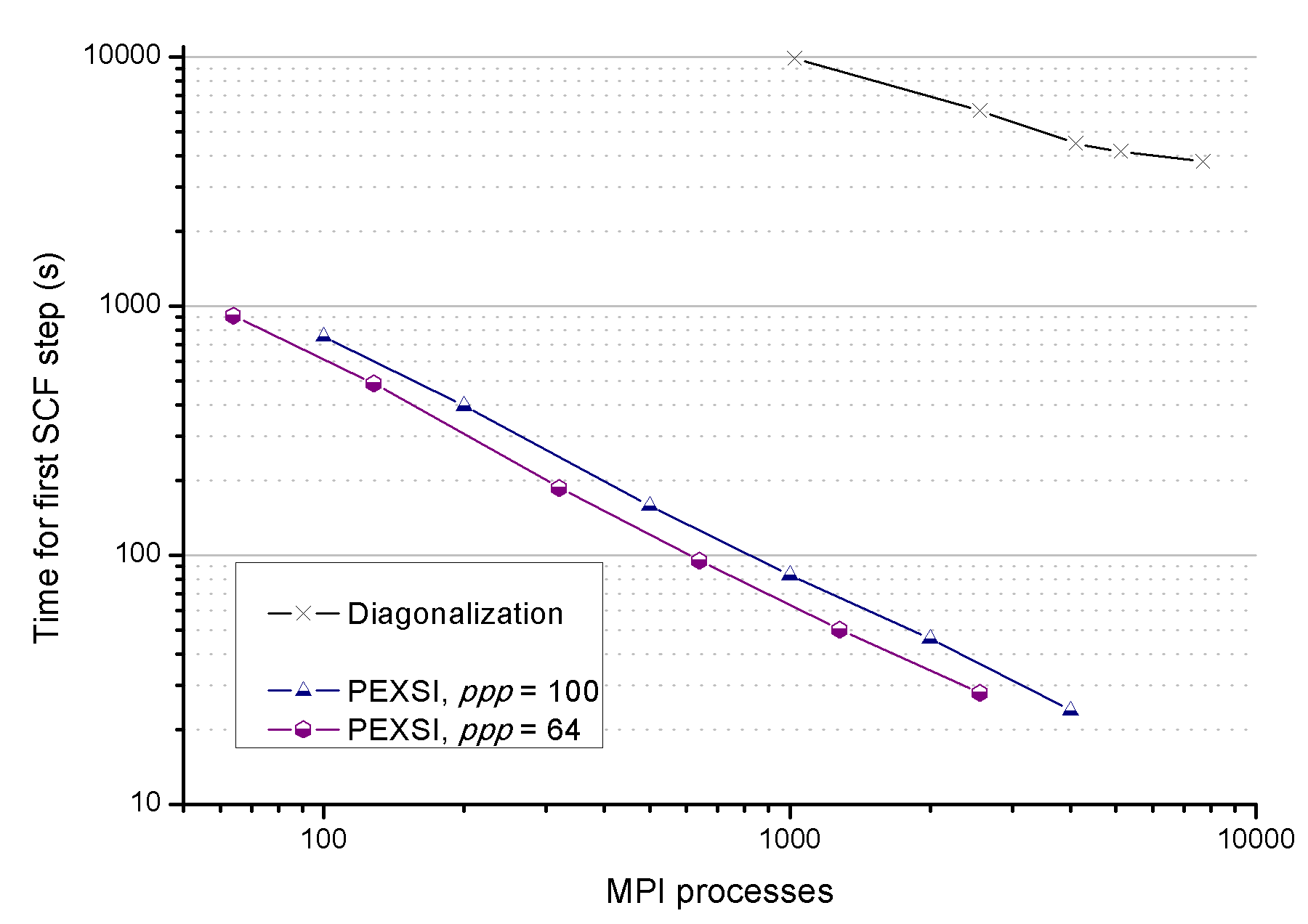}}\\
  \subfloat[C-BN$_{0.00}$]{\includegraphics[width=0.8\linewidth]{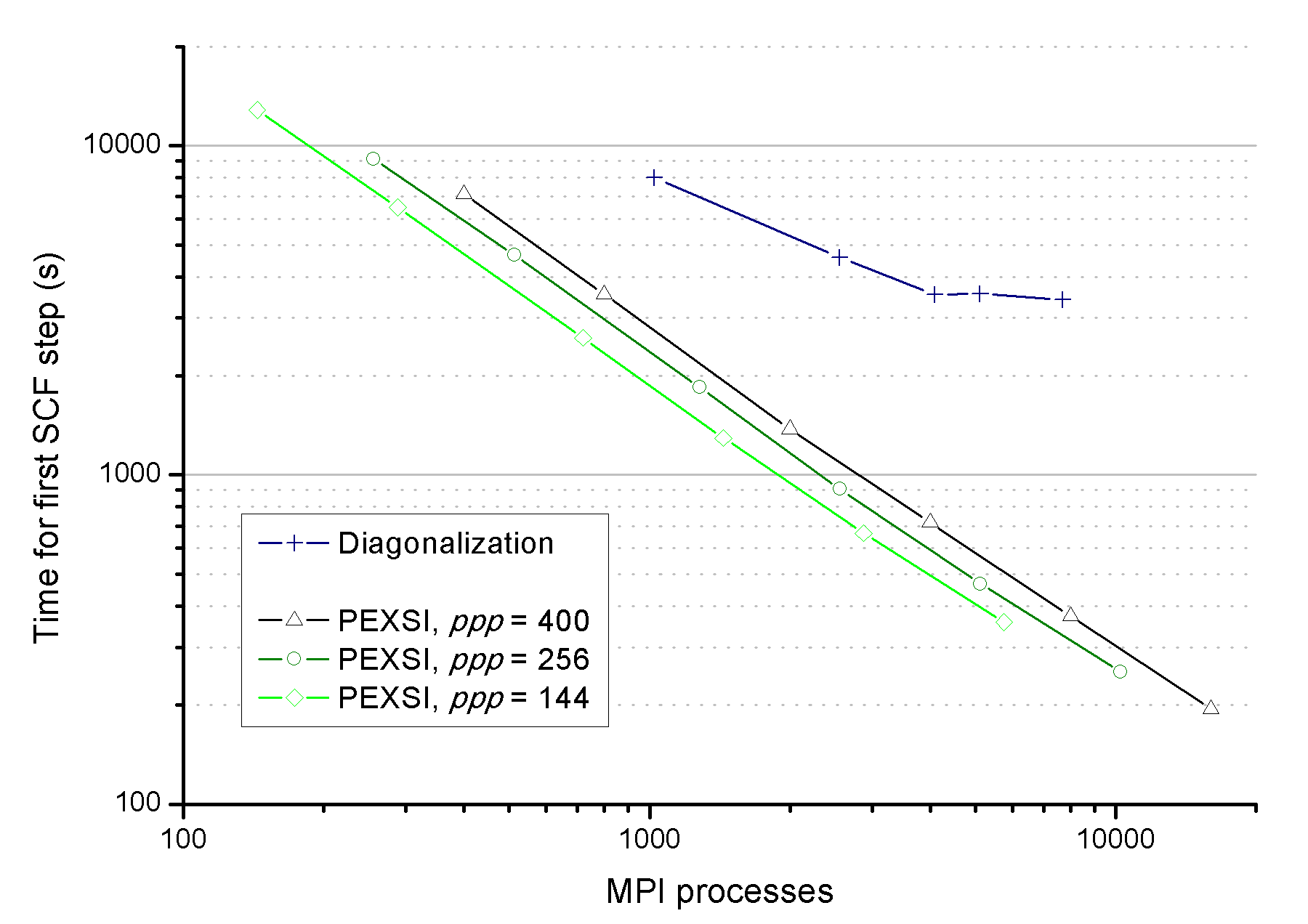}}\\
  \subfloat[H$_2$O-125]{\includegraphics[width=0.8\linewidth]{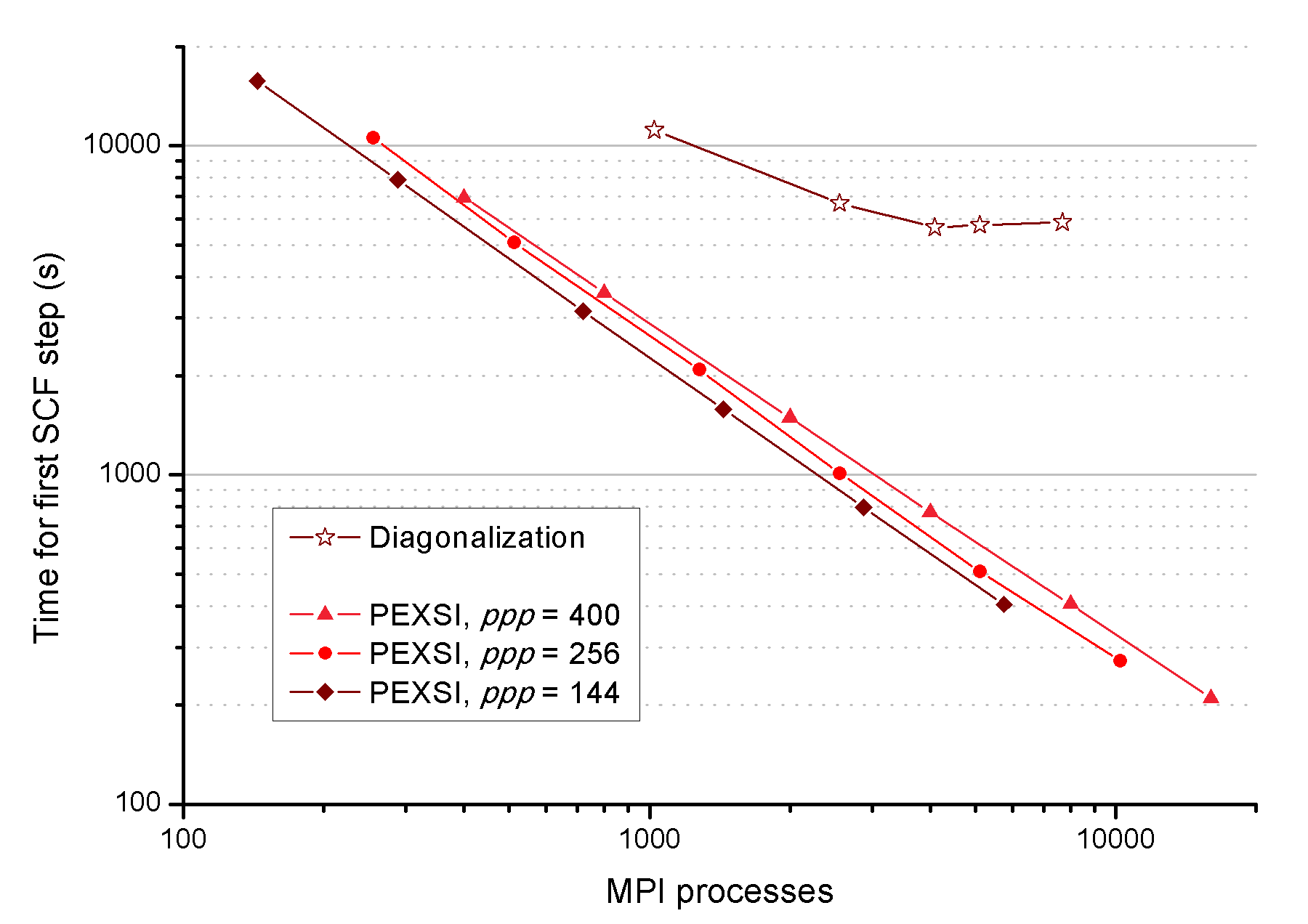}}
  \caption{(color online) Parallel strong scaling of SIESTA-PEXSI
           and the diagonalization approach when they are applied to
           DNA-25, C-BN$_{0.00}$ and H$_2$O-125 problems.}
  \label{fig:efficiencyStrongScaling}
\end{figure}

In Fig.~\ref{fig:efficiencyStrongScaling}, we plot the wallclock
time required to complete the first SCF step as a function of the
total number of processors used, for both SIESTA-PEXSI and the
standard diagonalization method in SIESTA, when they are used to solve
the DNA-25, C-BN$_{0.00}$, and H$_2$O-125 problems.
In our experiments we used $P=40$ poles in the pole expansion
approximation of the Fermi-Dirac function and varying degrees of
concurrency over poles: $np = k \times ppp$, where $k \in \{1, 2, 5,
10, 20, 40\}$. When $k=1$, we simply loop serially over poles and
perform parallel selected inversion on $ppp$ processors at each pole.
When $k = P =40$, full concurrency is achieved.  For each test
problem we connect all measurements that correspond to the same $ppp$
with a line. The nearly perfect scaling exhibited by these lines
reflects the embarrassingly parallel nature of pole expansion.
The further reduction in the wallclock time upon increasing $ppp$
depends for a given system on its size and the degree of sparsity
of the Cholesky factor $L$ (or in $L$ and $U$ factors in the $LU$
factorization).
We observe that increasing $ppp$ from 64 to 100 leads to additional
reduction in wallclock time by a factor of 1.2 for the DNA-25 problem.
The $L$ matrix of DNA is very sparse.
The $L$ matrices of C-BN and H$_2$O are less sparse, and more
processors can be effectively used to reduce the wallclock time of
selected inversion. We observe that increasing $ppp$ from 144 to 400
leads to an additional speedup of 1.8 for C-BN$_{0.00}$, and of 2 for
H$_2$O-125.

%
In Fig.~\ref{fig:efficiencyStrongScaling}, we also plot the
wallclock time used by the standard diagonalization-based
procedure. The computational complexity of diagonalization does not
depend on the sparsity of the $H$ and $S$ matrices but only on their
size, which is similar for all three systems, and therefore the performance of
the method on these problems is comparable. The diagonalization curves
in Fig.~\ref{fig:efficiencyStrongScaling} start at around 1000
processors because this is the minimum number that would allow these
problems to fit in memory, as the dense form of $H$ and $S$ is needed
by the algorithm. A reasonably good parallel scaling can be observed
up to 4000 processors, but the performance degrades after that.

We can also see that when around 1000 processors are used in
the computation, SIESTA-PEXSI is approximately one order of magnitude
faster than the diagonalization method for the C-BN$_{0.0}$ and
H$_2$O-125 problems, and approximately two orders of magnitude faster
for the DNA-25 problem. The performance gap between SIESTA-PEXSI and
diagonalization widens as the number of processors used in the computation
increases.  This is due to the relatively limited scalability of
ScaLAPACK, and that PEXSI can more efficiently utilize a large number of
cores thanks to the two-level parallelism.

Even though there is a clear advantage of SIESTA-PEXSI in being able
to use large numbers of processors, we should point out that its
smaller memory footprint means that it can operate also with relatively
small numbers of processors. Thus on Edison we can solve the DNA-25
and C-BN$_{0.00}$ problems by SIESTA-PEXSI with as few as 144
processors for C-BN$_{0.00}$ and 64 processors for DNA-25.  For the
DNA-25 problem, running SIESTA-PEXSI on 64 processors is still more
than four times faster than running the standard diagonalization
procedure on 5120 processors.

To measure the weak scaling of SIESTA-PEXSI, and compare it with that
of the diagonalization-based procedure, we ran both solvers on
multiple instances of the DNA, C-BN and H$_2$O systems with different
sizes.  We adjust the number of processors so that it is approximately
proportional to the dimension of the $H$ and $S$ matrices, i.e., as
the problem size increases, we use more processors to solve the larger
problem.  For diagonalization, the number of processes for each system
size are chosen so that there are approximately 40-50 orbitals per
processor. 
For PEXSI, in all weak scaling tests, we use $P=40$ poles and fully
exploit the concurrency at the pole expansion level, i.e., the
selected inversions associated with different poles are carried out
simultaneously on different groups of $ppp$ processors, for all
sizes. This means that the increase in processor count with problem
size is achieved with progressively higher $ppp$ values.  

In Table~\ref{tab:exampleWeakScaling} we report the
wallclock time $t_{iter}$ used to complete the first SCF step, and use the
$t_{iter}$ for the smallest size in each DNA, C-BN and H$_2$O
series as the basis for measuring weak scaling.  If we increase the
number of processors by a factor of $\alpha$ when the problem size is
increased by a factor of $\beta$, the ideal weak scaling factor (in
terms of the wall clock time) is
\[
s_w = \frac{\beta}{\alpha},
\]
for DNA problems due to the quasi-1D nature.  The ideal weak
scaling factors are 
$\frac{\beta^{3/2}}{\alpha}$ and $\frac{\beta^2}{\alpha}$
for C-BN and H$_2$O systems, due to the quasi-2D and 3D nature of the 
systems, respectively. Table~\ref{tab:exampleWeakScaling} shows the 
ideal time for the PEXSI runs computed using those ideal weak scaling
factors for each instance $t_{ideal} = t_0\times s_w$, where $t_0$ is
the wallclock time for the smallest instance in each series. The
weak-scaling for diagonalization is close to the expected cubic scaling
$s_w = \frac{\beta^3}{\alpha}$, and the corresponding ideal time is not
shown for simplicity.

\begin{table}[htb]
\begin{center}
  \begin{tabular}{@{\extracolsep{3pt}}|l|rr|rrr|}
    \hline
            & \multicolumn{2}{c|}{Diagonalization} &
    \multicolumn{3}{c|}{Siesta-PEXSI} \\
    Example       &  Proc. & $t_{iter}$ (s)& Proc.& $t_{iter}$ (s)&
    $t_{ideal}$ (s)\\
    \hline
    DNA-1         &    128 &    7.1 &  360  &   3.8  &    3.8 \\
    DNA-4         &    512 &  123   & 1000  &   6.8  &    5.5 \\
    DNA-9         &   1280 &  606   & 1960  &  10.9  &    6.3 \\
    DNA-16        &   2560 & 2005   & 2560  &  16.6  &    8.6 \\
    DNA-25        &   4096 & 4118   & 4000  &  24.4  &    8.6 \\
    \hline
    C-BN$_{2.3}$  &    720 &   96.5 &  1440 &   69.2   &   69.2   \\
    C-BN$_{1.43}$ &   1280 &  393   &  2560 &  136   &  105   \\
    C-BN$_{0.57}$ &   2560 & 1422   &  5680 &  188   &  140   \\
    C-BN$_{0.00}$ &   4096 & 3529   & 10240 &  248   &  157   \\    
    \hline
    H$_2$O-8      &    256 &   16.8 &   640 &   12.3 &  12.3  \\
    H$_2$O-27     &    720 &  272   &  2560 &   47.3 &  35.0  \\
    H$_2$O-64     &   2048 &  1375  &  5760 &  126   &  87.5  \\
    H$_2$O-125    &   4096 &  5641  & 10240 &  314   & 188    \\
    \hline
  \end{tabular}
\end{center}
\caption{Configurations and times for the first SCF iteration, as
presented in  Fig.~\ref{fig:efficiencyWeakScaling}. }
\label{tab:exampleWeakScaling}
\end{table}

\begin{figure}[htb]
  \begin{center}
    \includegraphics[width=0.9\columnwidth]{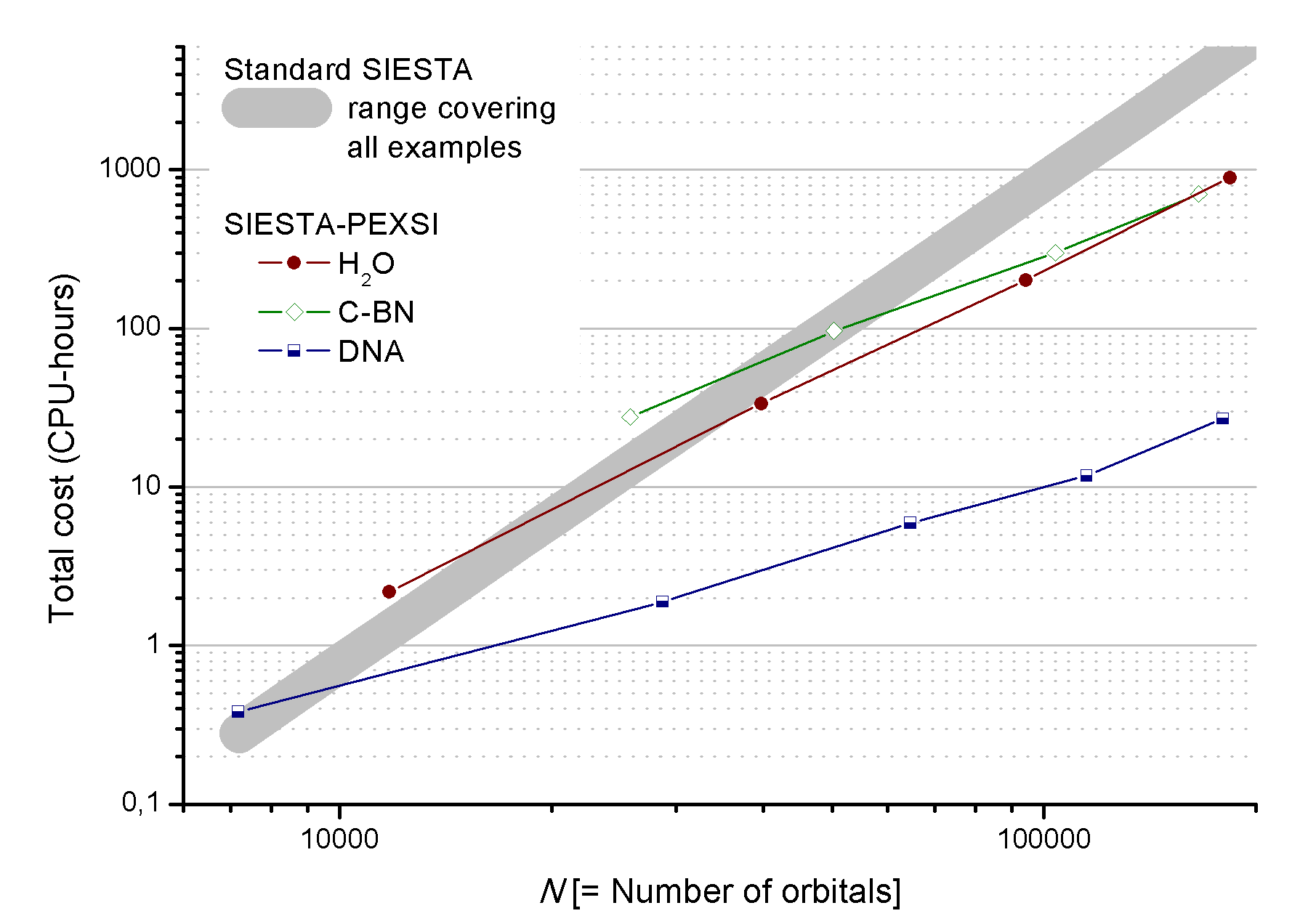}
  \end{center}
  \caption{(color online) Weak scaling based on computational cost (time $\times$
  number of processors) for the first SCF step for the H$_2$O, C-BN,
  and DNA examples.  The points correspond to the various problem
  sizes as listed in Table~\ref{tab:exampleFeatures}. The times for
  diagonalization show some variations due to differences in the
  construction of $H$ and the building of the $\Gamma$ matrix, and
  are represented by the gray stripe. The configurations and results
  can be found in more detail in Table
  \protect\ref{tab:exampleWeakScaling}. }
  \label{fig:efficiencyWeakScaling}
\end{figure}

Another way to study the weak scaling of the SIESTA-PEXSI approach
is to examine how the total computational cost, which is the product
of the wallclock time and the number of processors (i.e., the first
two columns for each method in Table~\ref{tab:exampleWeakScaling}),
change with respect to the problem size.  For perfect weak scaling, the
change in cost should match that predicted by the computational
complexity of the sequential algorithm.  We plot the cost involved in
solving each instance of the DNA, C-BN and H$_2$O problems in
Fig.~\ref{fig:efficiencyWeakScaling}, which shows that the data for
each series falls approximately on a line. 
The (fitted) slope of the cost for DNA, which
includes both factorization and selected inversion, is observed to be
approximately 1.3, higher than the $O(N)$ ideal scaling for quasi-1D
systems.  
For C-BN the slope of the fitting line is approximately 1.7, which is
a bit larger than the $O(N^{1.5})$ ideal scaling for quasi-2D
systems.
The slope of the line for the H$_2$O series is approximately 2.18, again
slightly larger than the expected $O(N^{2})$ scaling for 3D systems.
The observed degradation in parallel scaling (relatively larger for DNA) 
comes about because
neither selected inversion nor factorization (which has the same
asymptotic complexity) can scale perfectly due to the complicated data
communication and task dependency involved. 

\begin{figure}[htb]
  \centering
  \includegraphics[width=\columnwidth]{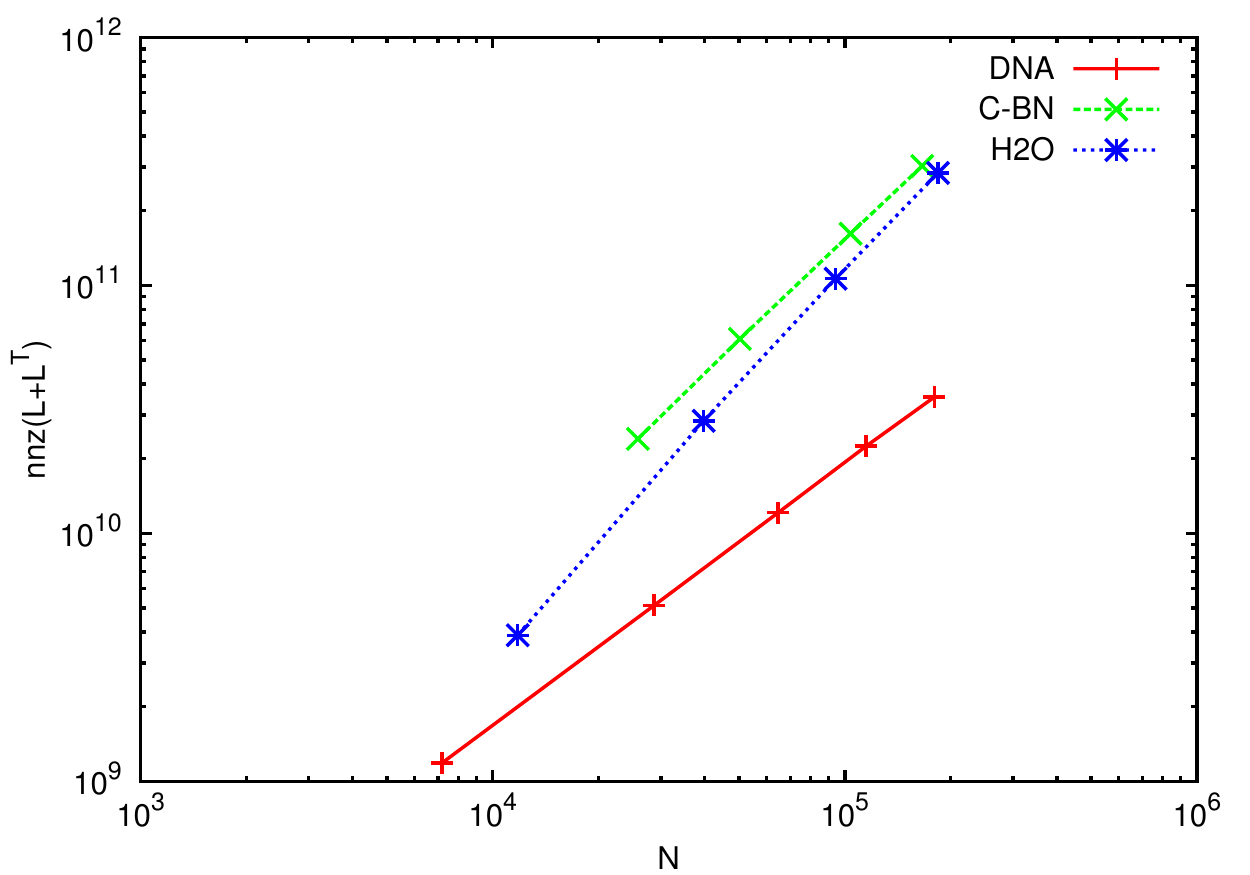}
  \caption{(color online) Number of non-zeros in the Cholesky factor of $H$ for the
    systems shown on Fig.~\protect\ref{fig:efficiencyWeakScaling}, as
    a function of the dimension of $H$. }
  \label{fig:cholesky_nnz}
\end{figure}

It should be noted also that the ideal weak-scaling data in Table 
\ref{tab:exampleWeakScaling} depends on the size and
processor count chosen for the smallest instance of each series. We
have already mentioned that all systems considered in our timings have
been run with complete parallelization over poles. This limits the
gains from increasing processor counts to the actual efficiency of
selected-inversion and factorization within a pole, as a function of
$ppp$. 
Roughly, the
computational load for a given instance will depend on the number of
non-zeros in the Cholesky factor $L$, which is plotted as a function
of matrix size in Fig.~\ref{fig:cholesky_nnz}. 
It can be seen that
C-BN has the highest such non-zero density, closely followed by
H$_2$O. This is because C-BN has a more densely packed
structure than H$_{2}$O, and thus a larger prefactor,
but there is a clear trend for a faster increase in the number of non-zeros
in the Cholesky factor for H$_{2}$O, 
in agreement with the asymptotic complexity
of 3D and quasi-2D systems, respectively. 
The DNA system has significantly fewer non-zero values in L for a
given size of H, and so it can use a relatively smaller number
of processors per pole efficiently.


%


We can also see from Fig.~\ref{fig:efficiencyWeakScaling} that the
crossover point at which SIESTA-PEXSI becomes more efficient (in terms
of cost) than the standard diagonalization procedure in SIESTA is
around matrix dimension 
$N$=7000 ($\sim$700 atoms) for the DNA problem, $N$=50,000 ($\sim$4,000 atoms) 
for the C-BN problem and $N$=40,000 ($\sim$5,000 atoms) for the H$_2$O problem. 
The actual orderings of the crossover
points depend on the dimensionality (through the power of N in
the scaling), and also on a prefactor which is system dependent.

Since more processors can be used efficiently, the advantage of
SIESTA-PEXSI in time-to-solution performance is even more pronounced
than the benefit from the reduction of total work load, and is already evident for smaller systems, as
can be seen in Table \ref{tab:exampleWeakScaling}.

The conclusions regarding the performance comparison between PEXSI
and diagonalization, being based on the asymptotic scaling of the
algorithm, remain largely valid even taking into account the efficiency
improvements that can been obtained by refactoring some of the internals
of ScaLAPACK~\cite{Auckenthaler2011}.

\subsection{The search for chemical potential}\label{subsec:robustmu}

As we discussed in section~\ref{subsec:chemicalpotential}, the current
implementation of SIESTA-PEXSI uses a combination of an
inertia counting bracketing technique and Newton's method to determine
the chemical potential $\mu$ that satisfies the nonlinear equation
$N_\beta(\mu) = N_e$, where $N_\beta(\mu)$ is defined in
\eqref{eqn:Nmu}.  In this section, we demonstrate the effectiveness of
this approach for both metallic and insulating systems.

\begin{figure}[ht]
  \begin{center}
    \includegraphics[width=\linewidth]{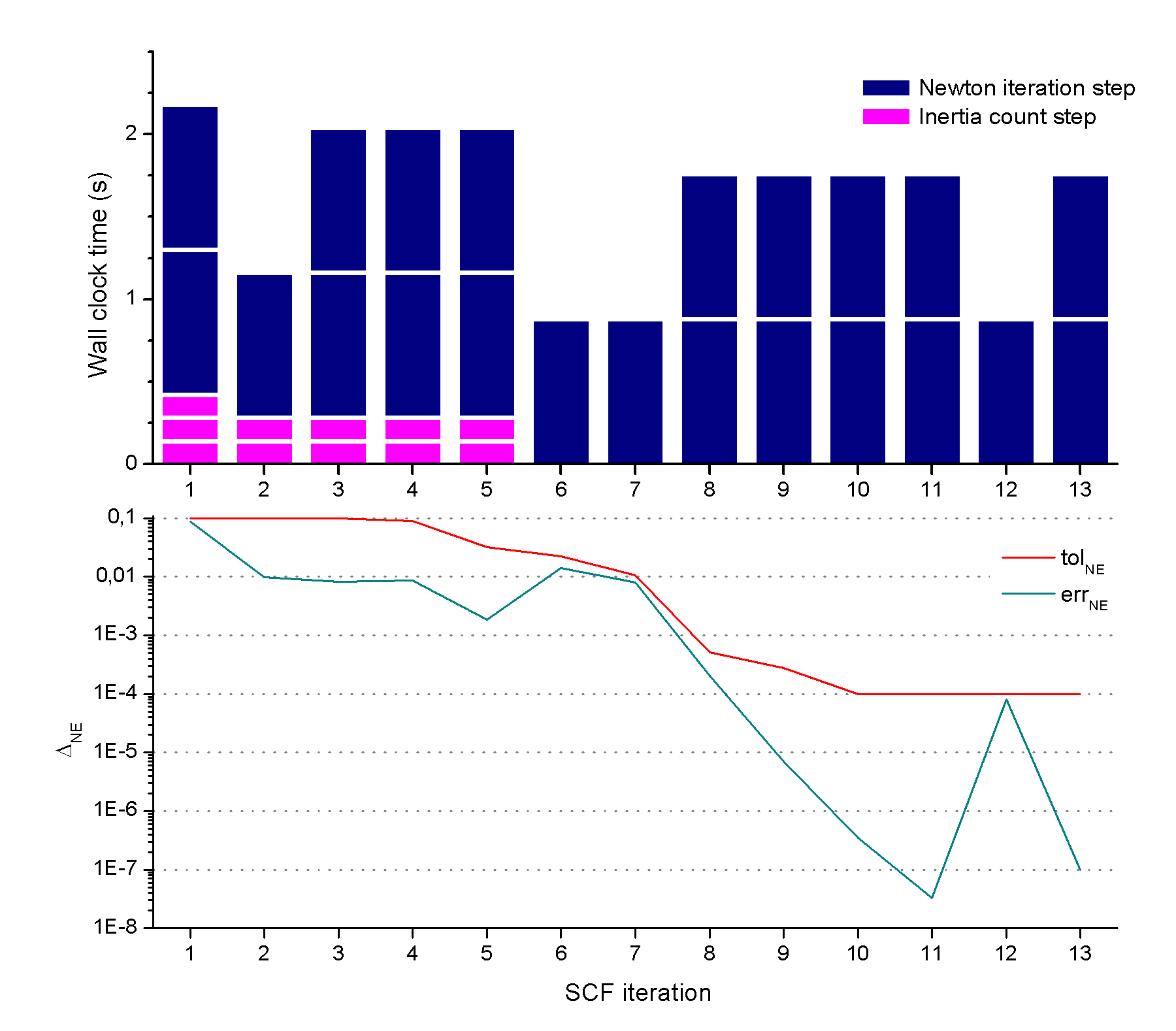}
  \end{center}
  \caption{(color online) Convergence history in the SCF cycle for the SiH system. 
    The top graph shows
    the time per SCF step and a breakdown into the number of inertia
    counting and PEXSI solver iteration (Newton iteration) steps. The bottom diagram
    draws the (adaptive) target tolerance and the actual error in the number of
    electrons at the end of each SCF step.}
  \label{fig:convSiH}
\end{figure}

The metallic example is the SiH system with $257$ electrons.
Since our calculation is spin-restricted, the odd number of electrons
is solely controlled through the fractional occupation of the
H-derived state in the gap. Any small perturbation of the chemical
potential can change the number of electrons, and the perturbation is
temperature dependent (we use 300 K).  Despite the relatively small
system size, this can be considered as the most difficult case for
finding the chemical potential without having access to the
eigenvalues.  Fig.~\ref{fig:convSiH} illustrates the number of
inertia counting cycles and PEXSI solver steps needed 
to obtain a chemical potential
that satisfies $|N_\beta(\mu) - N_e | < 10^{-4}$ for the converged SiH
system when using 60 poles.  The calculation starts from a wide
initial $\mu$ search interval of $(-27.0, 0.0)$ eV, and uses an
adaptive electron tolerance which starts at a coarse $tol_{Ne}$=0.1
and progressively tightens towards the target value of $10^{-4}$ as
the deviation from self-consistency in the H matrix elements moves
towards its tolerance target of $10^{-5}$ Ry. The
number of inertia counting steps needed to narrow down the interval
that contains the true chemical potential is roughly $2\sim 3$ in the
first few SCF steps. 
After the $5$th iteration, the inertia counting procedure is turned off.
The number of PEXSI solver steps needed is at most $3$ in the first
few SCF steps, and it becomes 2 after the $6$th SCF step.
Fig.~\ref{fig:convSiH} also shows the absolute error
$|N_\beta(\mu)-N_e|$ at each SCF step, which is always below the
set tolerance.


\begin{figure}[h]
  \begin{center}
    \includegraphics[width=\linewidth]{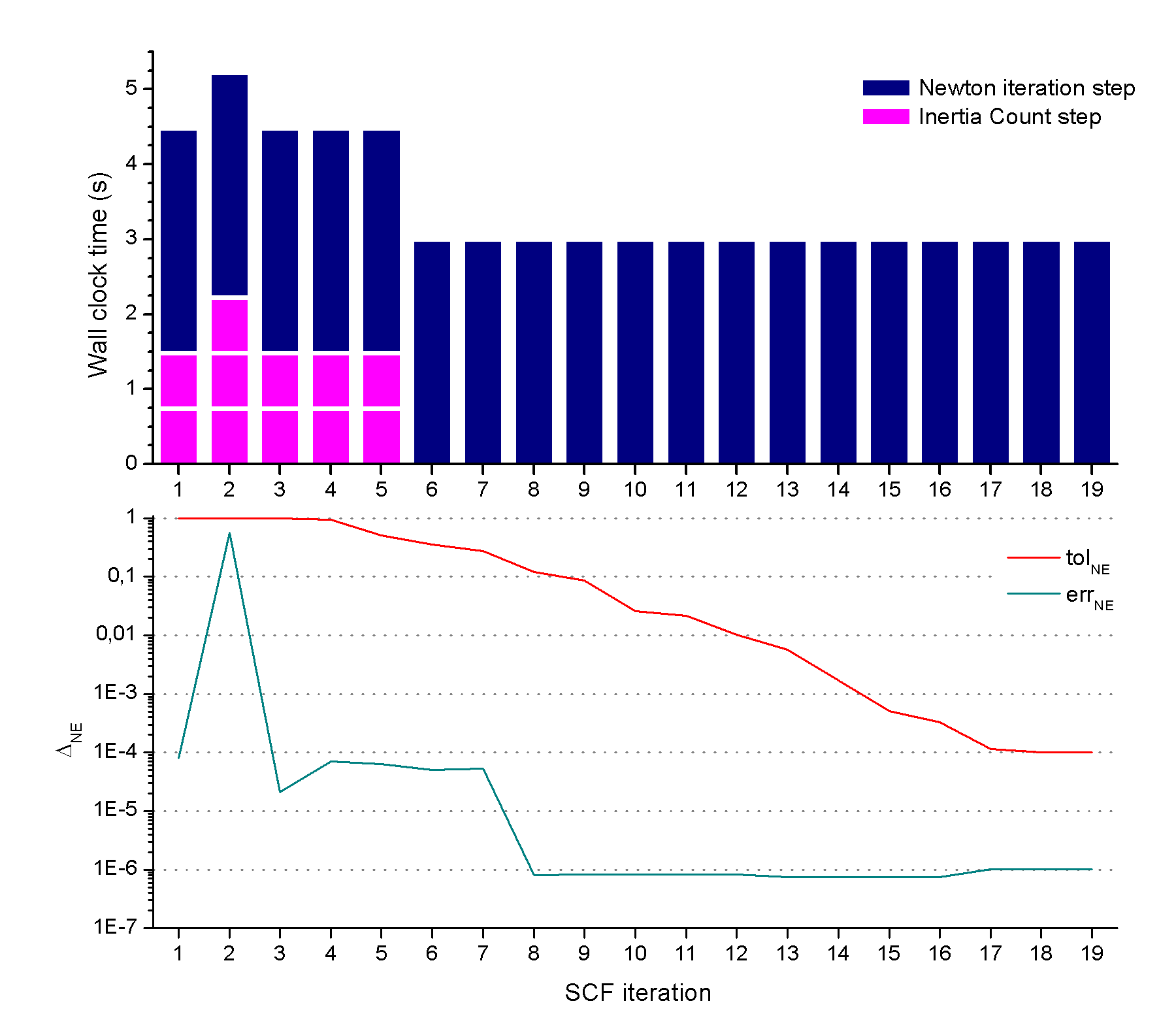}
  \end{center}
  \caption{(color online) Convergence history in the SCF cycle for the DNA-1 system.
    The top graph shows
    the time per SCF step and a breakdown into the number of inertia
    counting and PEXSI solver iteration (Newton iteration) steps. The bottom diagram
    draws the (adaptive) target tolerance and the actual error in the number of
    electrons at the end of each SCF step.}
  \label{fig:convDNA1}
\end{figure}

Our test for insulating systems is the DNA-1 system with $2442$
electrons, at $300$ K and using 50 poles.  
Fig.~\ref{fig:convDNA1} illustrates the
number of inertia counting and PEXSI solver steps needed to reduce
$|N_\beta(\mu) - N_e|$ to a level below a final tolerance of
$10^{-4}$, starting at a coarser tolerance of $1$.
The calculation also starts from a wide initial interval of
$(-27.0, 0.0)$ eV. Usually two inertia counting steps are needed 
in the first few SCF iterations. The second iteration requires 
one more due to a large jump in H.
After the $5$th iteration, the inertia counting
procedure is completely turned off. Throughout the SCF cycle, which ends
when the deviation from self-consistency in H is below $10^{-5}$ Ry,
just one PEXSI solver call is needed to provide a solution with a
number of electrons within the tolerance. This is because $\mu$ is
kept within the gap by the bracketing procedure.

\subsection{DOS and LDOS}

\begin{figure}[htb]
  \centering
  \subfloat[]{\includegraphics[width=0.49\linewidth]{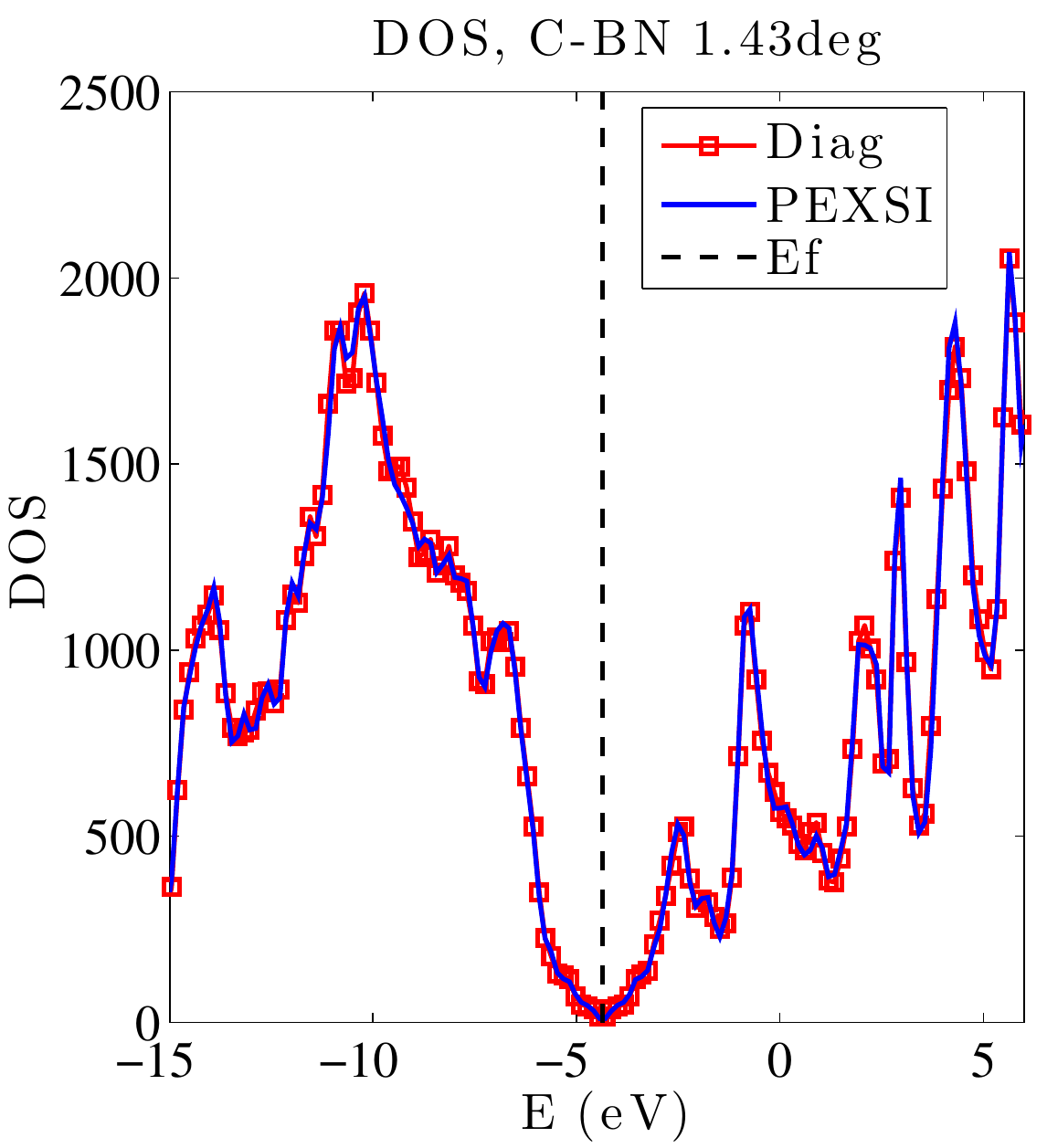}}
  \subfloat[]{\includegraphics[width=0.49\linewidth]{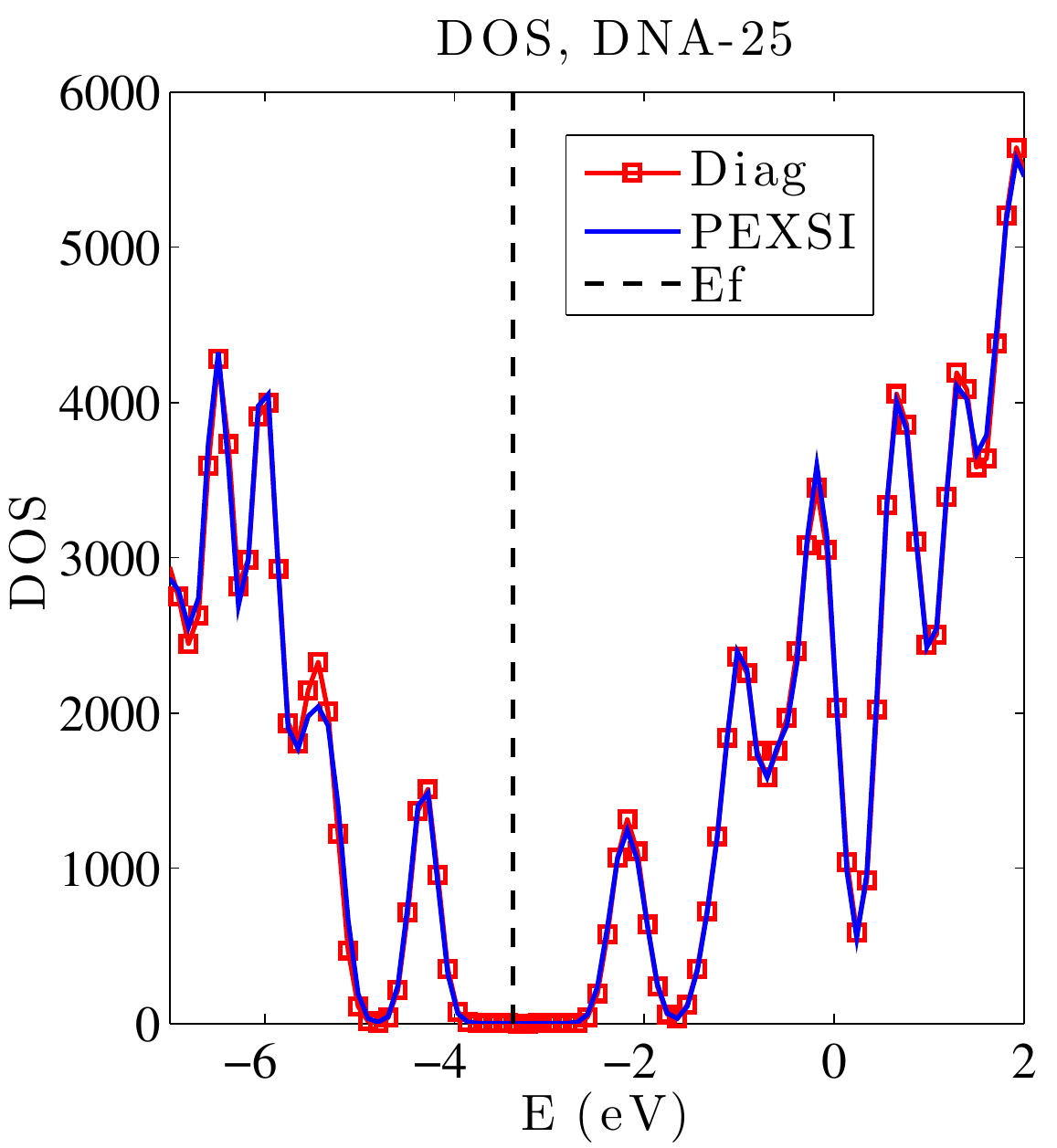}} 
  \caption{(color online) DOS around the Fermi level for the 1.43 degree C-BN and
    DNA-25 systems.}
  \label{fig:DOS}
\end{figure}

\begin{figure}[htb]
  \centering
      \includegraphics[width=0.6\columnwidth]{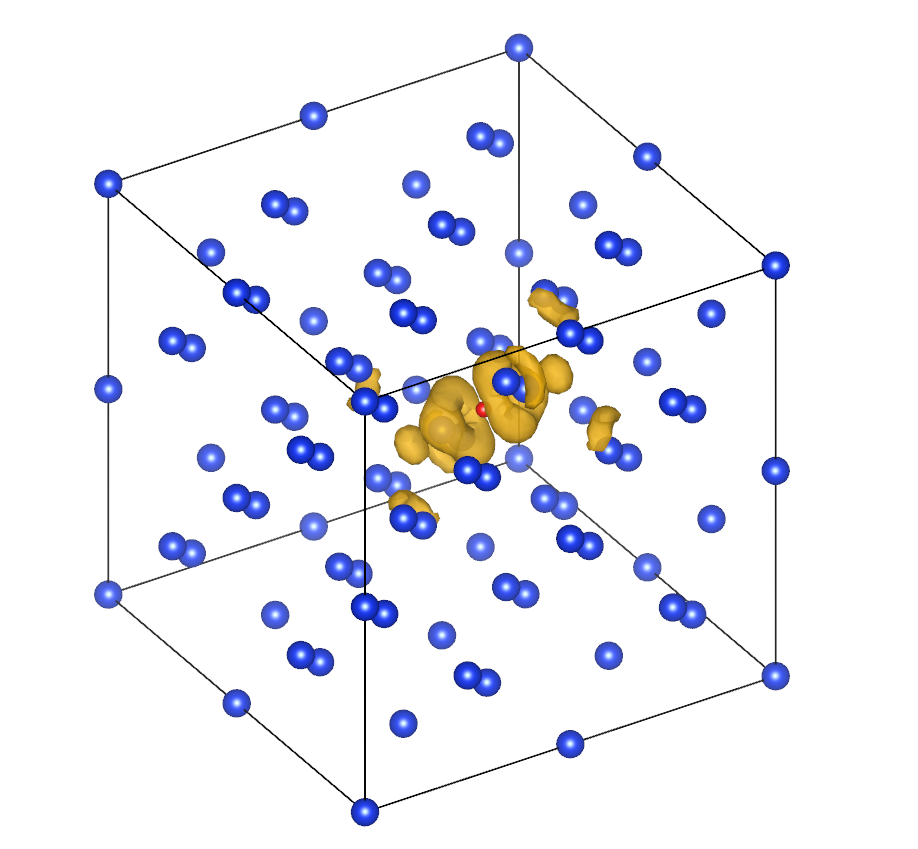}
  \caption{(color online) Isosurface of the LDOS around the
    Fermi level for the interstitial H atom in the Si
    crystal. }
  \label{fig:LDOS}
\end{figure}

In section \ref{subsec:dos}, we described how to use SIESTA-PEXSI to
obtain spectral information of the atomistic system such as DOS and LDOS 
without computing eigenvalues and eigenvectors of $(H,S)$.  
In Fig.~\ref{fig:DOS} (a) we plot the DOS obtained from the 
standard diagonalization method and the inertia counting procedure
implemented in SIESTA-PEXSI for the C-BN$_{1.43}$ system 
near the Fermi level (``E$_f$'').  
We observe nearly perfect agreement between the DOS curves
obtained from these two approaches. 
Fig.~\ref{fig:DOS} (b) shows the DOS near the Fermi level for the DNA-25
system using both the inertia counting method and diagonalization.
The usage of PEXSI for evaluating DOS
without obtaining eigenvalues is also significant for systems at large
size.  For the DNA-25 system, using 64 $ppp$ and 3200 processors to
evaluate $200$ points in the DOS near the Fermi energy (each group of
64 processors evaluate $4$ points of cumulative DOS with inertia
counting) only takes
$34$ sec, while diagonalization using the same number of processors
takes $4865$ sec.  It should be noted that the performance of PEXSI can
further be improved by simply using more cores: after the SCF converges,
one can restart the calculation to compute the DOS using the converged
density matrix, and use $12800$ processors to compute $200$ points in
the DOS in parallel.  In such case, the wall clock can be reduced to $9$
sec.
%
Fig.~\ref{fig:LDOS} presents the LDOS for the SiH system for an energy
interval of width 0.4 eV around the Fermi level, showing the state due
to the added H atom in the bulk Si system.

\section{Conclusion}\label{sec:conclusion}

We have combined the pole expansion and selected inversion (PEXSI)
technique with the SIESTA method for Kohn-Sham density functional theory
(KDSFT) calculation.  The resulting SIESTA-PEXSI method can efficiently
use more than $10,000$ processors, and is particularly suitable for
performing large scale \textit{ab initio} materials simulation on high
performance parallel computers. The SIESTA-PEXSI method does not
compute eigenvalues or eigenvectors of the Kohn-Sham
Hamiltonian, and its accuracy is fully comparable to that obtained from
the standard matrix diagonalization based SIESTA calculation for 
general systems, including insulating, semi-metallic, and metallic 
systems at low temperature.

The current implementation of SIESTA-PEXSI does not yet support spin-polarized 
systems, but this can be achieved with minor changes to the code.
Furthermore, the code does not support \textbf{k}-point sampling, as in
principle it is not needed for large-enough systems. However, in some
cases, as in graphene and similar semi-metallic systems, an
appropriate computation of the spectral properties such as the DOS does
need \textbf{k}-point sampling even when very large supercells are used. 
We will address this problem in future work, which requires modifying
PEXSI to handle complex Hermitian $H$ matrices. In such case the shifted
matrix $H-zS$ is no longer symmetric but is only structurally symmetric.
This further development will also be useful for \textit{ab initio}
study of non-collinear magnetism and spin-orbit coupling effects.


\section*{Acknowledgment}
This work was partially supported by the Laboratory Directed Research
and Development Program of Lawrence Berkeley National Laboratory under
the U.S.  Department of Energy contract number DE-AC02-05CH11231, by
Scientific Discovery through Advanced Computing (SciDAC) program funded
by U.S.  Department of Energy, Office of Science, Advanced Scientific
Computing Research and Basic Energy Sciences, by the Center for Applied
Mathematics for Energy Research Applications (CAMERA), which is a
partnership between Basic Energy Sciences and Advanced Scientific
Computing Research at the U.S Department of Energy (L. L.  and C. Y.), by the
European Community's Seventh Framework Programme [FP7/2007-2013] under
the PRACE Project grant agreement number 283493 (G. H.), and by the Spanish
MINECO through grants FIS2009-12721-C04-03, FIS2012-37549-C05-05 and
CSD2007-00050 (A. G.).

%

%

\end{document}